\documentclass[11pt]{article}
\usepackage{amsmath,amssymb,amsfonts,color,epsf}
\usepackage{slashed,latexsym}
\usepackage{graphics,graphicx,picture}
\usepackage{cite}
\makeatletter
\@addtoreset{equation}{section}
\makeatother

\newcommand{\hoch}[1]{$\, ^{#1}$}

\newcommand{\be}{\begin{equation}}
\newcommand{\ee}{\end{equation}}
\newcommand{\bea}{\setlength\arraycolsep{2pt} \begin{eqnarray}}
\newcommand{\eea}{\end{eqnarray}}

\usepackage[font=footnotesize,labelsep=newline,labelfont=sc,justification=centering,position=top]{caption}

\def\rmi{{\rm i}}
\newsavebox{\uuunit}
\sbox{\uuunit}
    {\setlength{\unitlength}{0.825em}
     \begin{picture}(0.6,0.7)
        \thinlines
        \put(0,0){\line(1,0){0.5}}
        \put(0.15,0){\line(0,1){0.7}}
        \put(0.35,0){\line(0,1){0.8}}
       \multiput(0.3,0.8)(-0.04,-0.02){12}{\rule{0.5pt}{0.5pt}}
     \end {picture}}

\def\be{\begin{equation}}
\def\ee{\end{equation}}
\def\ba{\begin{array}}
\def\ea{\end{array}}
\def\bea{\begin{eqnarray}}
\def\eea{\end{eqnarray}}
\def\bd{\begin{displaymath}}
\def\ed{\end{displaymath}}

\usepackage{color}
\definecolor{MyDarkBlue}{rgb}{0.15,0.15,0.45}    \definecolor{MyGreen}{rgb}{0.15,0.45,0.45}
\definecolor{MyPurple}{rgb}{0.55,0.25,0.55}
\usepackage[linktocpage=true]{hyperref}
\hypersetup{colorlinks=true,citecolor=MyPurple,linkcolor=MyPurple,urlcolor=MyGreen}

\begin{document}

\begin{flushright}
\hfill{ \
\ \ \ \ UG-14-14  \\  ICCUB-14-052  \\  TUW-14-09}
\end{flushright}
\vskip 1.2cm

\begin{center}
{\Large \bf The Non-Relativistic Superparticle in a Curved Background}
\end{center}
\vspace{20pt}
\begin{center}
{\Large {\bf }}

\vspace{10pt}

{\Large Eric Bergshoeff\hoch{1},
Joaquim Gomis\hoch{2},
Marija Kova\v{c}evi\'c\hoch{1}, \\ Lorena Parra\hoch{1,3},
Jan Rosseel\hoch{4} and Thomas Zojer\hoch{1}
}

\vspace{10pt}

\hoch{1} {\it Van Swinderen Institute for Particle Physics and Gravity, University of Groningen,\\
Nijenborgh 4, 9747 AG Groningen, The Netherlands}\\

\hoch{2} {\it Departament d'Estructura i Constituents de la Mat\`eria and Institut de Ci\`encies del\\
Cosmos, Universitat de Barcelona, Diagonal 645, 08028 Barcelona, Spain}

\hoch{3} {\it Instituto de Ciencias Nucleares, Universidad Nacional Aut\'onoma de M\'exico,\\
Apartado Postal 70-543, 04510 M\'exico, D.F., M\'exico}\\

\hoch{4} {\it Institute for Theoretical Physics, Vienna University of Technology,\\
                               Wiedner Hauptstr.~8--10/136, A-1040 Vienna, Austria}

\vspace{10pt}

\texttt{e.a.bergshoeff@rug.nl, joaquim.gomis@gmail.com, m.kovacevic@rug.nl,\\
        l.parra.rodriguez@rug.nl, rosseelj@gmail.com, t.zojer@rug.nl}

\vspace{30pt}
% \underline{ABSTRACT}
\end{center}

Using a component formulation, we construct the supersymmetric action for a superparticle in a
three-dimensional Newton--Cartan supergravity background and clarify its symmetries. Our construction proceeds by
first constructing the superparticle in a flat background. Next, by boosting up the background symmetries,
we replace in a first step the flat background by a Galilean supergravity background. In a second step the
Galilean supergravity  background is replaced by a Newton--Cartan supergravity background. We extend our results
by adding a supersymmetric cosmological constant and compare the non-relativistic superparticle with the relativistic
$\kappa$-symmetric three-dimensional superparticle.

\vspace{15pt}

\thispagestyle{empty}

\vspace{15pt}

 \vfill

\thispagestyle{empty}
\voffset=-40pt

\newpage

\tableofcontents

%\addtocontents{toc}{\protect\setcounter{tocdepth}{2}}

\newpage

%%%%%%%%%%%%%%%%%%%%%%%%%%%%%%%%%%%%%%%%%%%%%%%%%%%%%%%%%%%%%%%%%%%%%%%%%%%%%%%

\section{Introduction}

Over the past few years there has been an increased interest in extending the holographic principle to
non-relativistic theories. The main motivation is to widen the applications of
the conjectured anti-de Sitter/conformal field theory (AdS/CFT) correspondence and to test if it also holds
away from its original relativistic setting. There are reasons to expect that non-relativistic holography
has applications in effective descriptions of strongly correlated condensed matter systems \cite{Nishida:2007pj,
Son:2008ye,Balasubramanian:2008dm,Herzog:2008wg,Kachru:2008yh}; for a review, see e.g.~\cite{Hartnoll:2009sz}.
Usually, one only considers a non-relativistic setting at the boundary but one might also consider
non-relativistic models both in the bulk and at the boundary \cite{Bagchi:2009my}. In the context of the AdS/CFT
correspondence non-relativistic theories also occur as boundary geometries. For example, the authors of
\cite{Christensen:2013lma,Christensen:2013rfa} have found that gravity theories supporting Lifshitz
geometries lead to a boundary geometry that is characterized by a so-called torsional Newton--Cartan geometry.\footnote{
The torsion occurs due to the fact that the boundary geometry is invariant under an anisotropic scaling symmetry.}
More recently, non-relativistic gravity and Newton--Cartan structures, see e.g.~\cite{Julia:1994bs,Duval:2009vt}, have also been
used as a geometric way of realizing non-relativistic symmetries of effective field theories \cite{Son:2013rqa}.
Originally, before all these  developments took place, non-relativistic superstring theories and superbranes were
studied as special points in the parameter space of M-theory with non-relativistic symmetries
\cite{Gomis:2000bd,Danielsson:2000gi}. Non-relativistic strings were also thought as a possible
soluble sector within string theory or M-theory \cite{Gomis:2004pw,Gomis:2005pg}. This latter expectation was based on a
similar experience with the pp-wave/BMN limit \cite{Berenstein:2002jq}.

The goal of this paper is to make a first step in improving our knowledge on non-relativistic gravity and
particle/string/brane theories that underly a holography with a non-relativistic setting at both the boundary
and the bulk. In this work we will  consider supersymmetric theories and we will
restrict to  particles, or more precisely superparticles \cite{Casalbuoni:1976tz,Brink:1981nb}, only.

Non-relativistic $\mathcal{N}=2$ massive superparticles in ten dimensions in a flat background have already been
studied in \cite{Gomis:2004pw}. In this paper we wish to extend this analysis and consider superparticles in
a {\sl curved} non-relativistic supergravity background.
It should be stressed that, independent of the relation with non-relativistic holography, there is not much
literature on non-relativistic supersymmetry;
however, see e.g.~\cite{Puzalowski:1978rv,Clark:1983ne,Gauntlett:1990xq,deAzcarraga:1991fa,Duval:1993hs,Bergman:1995zr,Bergman:1996bx}.
This in itself provides ample reason to investigate this topic.

Part of this paper consists of a review of known results on non-relativistic superparticles in
a {\sl flat} background. To the best of our knowledge all non-relativistic superparticle actions with background
fields, with or without a cosmological constant, are new.

In the construction of massive (super-)particle actions an important role is played by symmetries, both
global and local ones. In particular, it is well-known that relativistic massive superparticles have an infinitely-reducible gauge
symmetry, called $\kappa$-symmetry
\cite{deAzcarraga:1982dw,Siegel:1983hh}, that eliminates half of the fermions. In the non-relativistic setting this
$\kappa$-symmetry corresponds to a fermionic gauge shift symmetry, i.e.~a St\"uckelberg symmetry \cite{Gomis:2004pw}.

When discussing the symmetries of (super-)particles in a curved background it is important to
distinguish between `proper' and `sigma model' symmetries \cite{Moore:1984dc,Hull:1995gk}. In the case of proper symmetries
the background (super-)gravity fields only transform through their dependence on the embedding coordinates of the
(super-)particle. On the other hand, in the case of sigma model symmetries the
background fields have their own transformation rules. We will clarify how these sigma model transformations are
related to the transformations of the (super-)gravity fields when viewed as the components of a (super-)gravity
multiplet defined in the target space.

To explain our construction of  non-relativistic superparticles in a curved background, we will first consider the
bosonic case and take as our starting point  a single massive non-relativistic particle in a flat background.
The action of such a  particle is invariant under the (global) Galilei symmetries, hence the name `Galilean' particle.
We next partially gauge the spatial target space translations of the Galilean particle such that the constant parameter
of a spatial translation is promoted
to an arbitrary function of time. The resulting extended symmetries are sometimes called the `acceleration-extended'
Galilean symmetries. In order to achieve this partial gauging the particle must move in a curved Galilean background
that is characterized by the Newton potential $\Phi$ \cite{golufb}.\footnote{There is also a different way of
gauging the spatial translations where the background is given by a vector field rather than a scalar
potential \cite{DePietri:1994je,Banerjee:2014pya}.}
We will call this particle a `Curved Galilean'
particle. Finally, we perform a full gauging of the Galilean symmetries such that the parameter of spatial translations
becomes an arbitrary function of the spacetime coordinates and the full symmetries of the particle action are the
general coordinate transformations. Actually, to perform this gauging it is necessary to extend the Galilei symmetries
with an additional central charge transformation\footnote{In sections \ref{sec:superparticle} and \ref{sec:superCC},
where we discuss supersymmetric particle actions, we will restrict ourselves to $d=3$. In that case, it is known that
the Bargmann algebra admits a second central charge, see e.g.~\cite{LL}. We will however not consider it in this paper.}
\cite{DePietri:1994je,Lukierski:2007nh,Andringa:2010it}. The background is now promoted to a
Newton--Cartan gravity background that is characterized by a spacelike Vielbein $e_\mu{}^a$, a timelike Vielbein
$\tau_\mu$ and a central charge gauge field $m_\mu$. We will call the corresponding particle a `Newton--Cartan' (NC)
particle. In table \ref{table:1} we have summarized the three different situations described above.

\bigskip
\begin{table}[ht]
\begin{center}
{\small
\begin{tabular}{|c|c|c|c|c|}
\hline \rule[-1mm]{0mm}{6mm}
 background fields & particle $\Lambda=0$ &symmetries $\Lambda=0$& particle $\Lambda \ne 0$   & symmetries $\Lambda\ne 0$ \\[.1truecm]
\hline \rule[-1mm]{0mm}{6mm}

none & Galilean & Galilei & NH & NH\\[.1truecm]

 $\Phi$ & Curved Galilean & a.e.~Galilei & Curved NH & a.e.~Galilei \\[.1truecm]

 $e_\mu{}^a$, $\tau_\mu$, $m_\mu$ & NC & g.c.t. & NC NH &g.c.t. \\[.1truecm]

\hline

\end{tabular}
}
\end{center}

 \caption{\sl This table gives an overview of the different bosonic backgrounds used in this paper (first column).
          The second and third  columns indicate the names we use for a particle moving in such backgrounds plus the
          symmetries of the corresponding particle action in the case of a zero cosmological constant ($\Lambda=0$).
          The last two columns give the same information for the case of a non-zero cosmological constant ($\Lambda\neq0$).
          We have denoted the acceleration-extended Galilei symmetries with `a.e.~Galilei' while g.c.t.~stands for
          general coordinate transformations. We have furthermore used the abbreviations NC (Newton--Cartan) and
          NH (Newton--Hooke).
         }\label{table:1}
\end{table}

In this work we will extend  the superparticle actions considered so far in two directions. First of all, we will
deform the different backgrounds with a cosmological constant $\Lambda$. For a flat\footnote{Note that after adding
the cosmological constant  space is flat while space-time is curved, see e.g.~\cite{Aldrovandi:1998im}.} background the Galilei
symmetries are then deformed to so-called Newton--Hooke (NH) symmetries \cite{Bacry:1968zf,dubois}\,\footnote{For
earlier works on NH symmetries including their supersymmetric extension, see e.g.~\cite{Gomis:2005pg,Gao:2001sr,
Gibbons:2003rv,Brugues:2006yd,Sakaguchi:2006pg,Lukierski:2007ed,Gomis:2008jc,Galajinsky:2009bk}.}.
We will call the corresponding particle  a `Newton--Hooke' particle.
% Physically, this system is equivalent to the non-relativistic harmonic oscillator.
The cosmological extensions of the Curved Galilean and NC particle are called the Curved NH
and NC NH particle, respectively. It turns out that the symmetries corresponding to these
particle actions are the same as in the absence of a cosmological constant, see table \ref{table:1}.
The reason for this is that the Galilean
and NH symmetries only differ in specific spatial translations which all become part of the same
time-dependent translations (Curved Galilean and Curved NH particle) or spacetime-dependent translations
(NC and NC NH particle). The three different cases with $\Lambda\neq0$ are
indicated in the last two columns of table \ref{table:1}.

The second extension we will consider is the one from the bosonic particle to the superparticle. This requires
a supersymmetric extension of the gravity backgrounds in the first place. Since non-relativistic supergravity
multiplets to our knowledge have only been explicitly constructed in three dimensions, we will only consider
superparticles in a three-dimensional (3D) background. A supersymmetric version of the 3D Galilean and NC
backgrounds was recently constructed by gauging the Galilei, or better Bargmann, superalgebra \cite{Andringa:2013mma}.
We will make full use of the construction of \cite{Andringa:2013mma} which, in particular, explains how to switch between
different backgrounds, with different symmetries, by partial gauging or partial gauge-fixing. Our aim will be to
investigate the action of a 3D superparticle first in a flat background and, next, in a Galilean and NC supergravity
background with and without a cosmological constant. To indicate the different cases we will use the same nomenclature
as in the bosonic case but with the word  particle replaced by superparticle.

\begin{figure}
\begin{center}
\setlength{\unitlength}{.7mm}
\begin{picture}(100,80)(-15,-15)
\thicklines
%
%\color{tascarletred}
\put(0,0){\circle{3}}            \put(-45,-2){Newton--Cartan}                    % NC
\put(0,30){\circle{3}}           \put(-45,28){Curved Galilean}
\put(0,60){\circle{3}}           \put(-27,58){Galilean}                         % Galilei
\put(70,0){\circle{3}}            \put(75,-2){NC NH}                    % NC
\put(70,30){\circle{3}}           \put(75,28){Curved NH}
\put(70,60){\circle{3}}           \put(75,58){Newton--Hooke}                     % Newton--Hooke
\put(0,2.5){\vector(0,1){25}} % gauge-fixing
\put(0,32.5){\vector(0,1){25}} % gauge-fixing
\put(70,2.5){\vector(0,1){25}} % gauge-fixing
\put(70,32.5){\vector(0,1){25}} % gauge-fixing
\put(-7,-10){$\Lambda=0$}
\put(63,-10){$\Lambda\neq0$}
\put(5,58){\ref{subsec:gal} \& \ref{sec:supergal}}
\put(5,28){\ref{subsec:new} \& \ref{sec:supernew}}
\put(5,-2){\ref{subsec:NC} \& \ref{sec:superNC}}
\put(43.1,58){\ref{subsec:CC} \& \ref{sec:superNH}}
\put(43,28){\ref{subsec:CC} \& \ref{subsec:superNHgal}}
\put(43,-2){\ref{subsec:CC} \& \ref{subsec:superNHNC}}
\end{picture}
\caption{\sl This figure displays the different  backgrounds used in this paper and the sections where they are discussed.
          For each background, the first section discusses the bosonic case while the second section
          treats the super\-sym\-metric case. The arrows indicate the direction of gauge-fixing.
          From left to right we turn on a cosmological constant $\Lambda$.}
\label{fig:backgrounds}
\end{center}
\end{figure}

\bigskip\noindent
 This work is organized as follows. In section
\ref{sec:bosonicparticle} we go through the known descriptions of the bosonic particle to introduce our notation
and discuss the different gaugings and gauge-fixings in a simple setting. After discussing the $\Lambda=0$ case
in some detail we repeat the analysis for a non-zero cosmological constant. Sections \ref{sec:superparticle} and
\ref{sec:superCC} are devoted to the supersymmetrization of the theories discussed in section \ref{sec:bosonicparticle}
without and with a non-zero cosmological constant, respectively. Section \ref{sec:kappa} is reserved for some
comments on the role of $\kappa$-symmetry in the non-relativistic case. A few technicalities, that are used in
the main text, are relegated to three appendices. Finally, we have given
 an alternative table of contents of this paper in figure \ref{fig:backgrounds} which summarizes in which
sections we discuss the different backgrounds that we will consider.

%%%%%%%%%%%%%%%%%%%%%%%%%%%%%%%%%%%%%%%%%%%%%%%%%%%%%%%%%%%%%%%%%%%%%%%%%%%%%%

\section{The Non-relativistic Bosonic Particle}\label{sec:bosonicparticle}

This section will serve as an introduction as well as a means to familiarize the reader with our notation.
Moreover, it is a useful warm-up exercise for the rest of the paper when we discuss the superparticle.
In subsection \ref{subsec:gal} we start by discussing the bosonic Galilean particle in a flat spacetime without gravitational dynamics.
Next, in subsection \ref{subsec:new} we will partially gauge the Galilei symmetries  to allow for arbitrary
time-dependent translations. This introduces the Curved Galilean particle which has gravitational
dynamics described by  a Newton potential, see \cite{Andringa:2012uz}. In subsection \ref{subsec:NC} we will perform
a further gauging and  obtain the Newton--Cartan (NC) particle action which is invariant under general coordinate
transformations. Finally, in subsection \ref{subsec:CC}, we repeat
the analysis of the previous three subsections in the presence of a cosmological constant.

\subsection{The Galilean Particle}\label{subsec:gal}

We  consider a bosonic particle moving in a flat $d$-dimensional background with
embedding coordinates $\{t(\tau)\,, x^i(\tau)\}\,, i=1,\dots,d-1,$ with $\tau$ the evolution parameter. The action
 describing the dynamics of this bosonic particle is given by
\begin{align}\label{NRbosonicparticle}
 S= \int d\tau\, \frac{m}{2}\, \frac{\dot x^i\dot x^i}{\dot t}\,.
\end{align}
The dot indicates a differentiation with respect to the evolution parameter $\tau$. The equations of motion for $t$
and $x^i$ corresponding to the action \eqref{NRbosonicparticle} are not independent. The first can be derived from
the latter which is given by
\begin{align}
 \frac{d}{d\tau}\left(\frac{\dot x^i}{\dot t}\right) &=0\,.
\end{align}
This implies the existence of a gauge transformation. In fact,
the action \eqref{NRbosonicparticle} is invariant under worldline reparameterizations with parameter $\rho(\tau)$:
\footnote{
In the following we refrain from denoting these worldline reparameterizations explicitly.}
\begin{equation}\label{wlrepar}
\delta t = \rho(\tau) \,{\dot t}\,,\hskip 2truecm \delta x^i = \rho(\tau) \,{\dot x}^i\,,
\end{equation}
together with the Galilei symmetries
\begin{equation}\label{Galtrans}
\delta t = -\zeta\,,\hskip 1.5truecm \delta x^i = \lambda^i{}_j x^j -v^i t -a^i\,.
\end{equation}
Here $(\zeta\,, \lambda^i{}_j\,, v^i\,, a^i)$ parametrize a (constant) time translation, spatial rotation, boost
transformation and space translation, respectively. Note that the Lagrangian \eqref{NRbosonicparticle} is invariant
under boosts only up to a total $\tau$-derivative. This has the consequence that the algebra of N\"other charges is
given by a centrally extended Galilei algebra which is called the Bargmann algebra \cite{levyleblond69,Marmo:1987rv}.
The transformations \eqref{Galtrans} are a representation of the Galilean algebra
\begin{eqnarray}\label{Galalg}
  &&\big[J_{ab},P_c \big] = -2\,\delta_{c[a}P_{b]} \,, \qquad\hskip .1truecm \big[G_a,H \big] = -P_a \,,\nonumber\\ [.1truecm]
                           && \big[J_{ab},G_c \big] = -2\,\delta_{c[a}G_{b]} \,,\qquad
                           \big[J_{ab}, J_{cd}\big] = 4 \delta_{[a[c}J_{d]b]}\,,
\end{eqnarray}
where the generators
\begin{align}
 \big\{H\,,J_{ab}\,,P_a\,,G_a\big\}
\end{align}
generate time translations, spatial rotations, space translations and boost transformations, respectively. The Bargmann
algebra has the additional commutation relation
\begin{align}\label{pqzcommutator}
 \big[P_a,G_b\big]=\delta_{ab}\,Z \,,
\end{align}
where $Z$ is the generator of central charge transformations. Physically, the occurrence of the central charge transformations
is related to the fact that at the non-relativistic level the mass of a particle is conserved.

\subsection{The Curved Galilean Particle}\label{subsec:new}

We next consider the Curved Galilean particle, i.e.~a bosonic particle in a Galilean gravity background
described by a Newton potential $\Phi(t,\vec x)$. The action for the Curved Galilean particle can be derived by gauging the
spatial translations \eqref{Galtrans} to allow for arbitrary time-dependent boost parameters $\xi^i(t)$. This
extension of the Galilean symmetries is called the `acceleration-extended Galilei symmetries' or the Milne symmetries
\cite{Duval:1993pe}.
As explained in \cite{Andringa:2012uz} this gauging procedure leads to the following action:
\begin{align}\label{bosonicgalileanparticle}
  S &= \int\, d\tau\, \frac{m}{2}\,\left[\frac{\dot x^i\dot x^i}{\dot t}  -2\dot t\, \Phi \right] \,.
\end{align}
The action \eqref{bosonicgalileanparticle} is  invariant under the worldline reparameterizations \eqref{wlrepar} and
under the acceleration-extended symmetries
\begin{align}\label{Newtrans}
\delta t = -\zeta\,,\hskip 1.5truecm \delta x^i = \lambda^i{}_j x^j -\xi^i(t)\,.
\end{align}
The acceleration-extended symmetries are not a proper symmetry of the action \eqref{bosonicgalileanparticle}.
Instead, the Newton potential should be viewed as a background field and
the acceleration-extended symmetries as sigma model symmetries. In particular, the transformation rule of the
background field, that we will denote by the symbol $\delta_{\mathrm{bg}}$, lacks the transport terms that are
present in the transformation rule associated to a proper symmetry, denoted in this paper by
$\delta_{\mathrm{pr}}$:\,\footnote{Assuming that
$x^\mu \ \rightarrow\ x^\mu+\delta x^\mu$ a transport term is given by $-\delta x^\mu\partial_\mu$ so that the second term in
\eqref{rel} cancels the transport term present in the proper transformation rule represented by the first term.}
\begin{align}\label{rel}
\delta_{{\rm bg}} = \delta_{{\rm pr}} + \delta x^\mu \partial_\mu\,.
\end{align}
Using this we find that the action \eqref{bosonicgalileanparticle} is invariant under the acceleration-extended symmetries
\eqref{Newtrans} provided the Newton potential $\Phi$ transforms as follows:
\begin{align}\label{bosonicsigma}
 \delta_{\text{bg}} \Phi = \frac{1}{\dot t}\frac{d}{d\tau}\left(\frac{\dot\xi^i}{\dot t}\right)x^i +\sigma(t)\,.
\end{align}
The second term with the arbitrary function $\sigma(t)$ represents a standard ambiguity
in any potential describing a force and gives a boundary term in the action \eqref{bosonicgalileanparticle}.
An important qualification of the background field is that it has to obey the constraint
\begin{align}\label{phieom}
 \partial^i\partial_i\Phi =0 \,.
\end{align}
We will  refer to this constraint as the equation of motion of the background field.

We can re-obtain the particle in flat space by setting
\begin{align}
 \Phi=0 \,.
\end{align}
This reduces the action and transformation rules of the Curved Galilean particle to the one of the  Galilean particle
given in the previous subsection.

\subsection{The Newton--Cartan Particle}\label{subsec:NC}

We now wish to extend the Curved Galilean particle  to a NC particle, i.e.~a  particle moving in a NC gravity
background and invariant under general coordinate transformations. To describe a  NC background we need  a temporal
Vielbein $\tau_\mu$ and  a spatial Vielbein $e_\mu{}^a\ , a=1,\dots, d-1,$. We furthermore need to introduce a central
charge gauge field $m_\mu$. The action for the NC particle was
already derived in \cite{Kuchar:1980tw}, see also \cite{Andringa:2012uz}, and is given by
\begin{align}\label{NCaction}
S = \int\, d\tau\,
\frac{m}{2}\,\left[\frac{\dot x^\mu e_\mu{}^a\,\dot x^\nu e_{\nu\,a}}{\dot x^\rho\tau_\rho} -2\,m_\mu\dot x^\mu\right]\,.
\end{align}
This action is invariant under the worldline reparameterizations \eqref{wlrepar} and under the gauged Galilei symmetries,
i.e.~general coordinate transformations. The embedding coordinates transform under these general coordinate transformations,
with parameters $\xi^\mu(x^\nu)$, in the standard way:
\begin{align}\label{xxi}
\delta x^\mu = -\xi^\mu(x^\nu)\,.
\end{align}
The transformation rules of the background fields $\tau_\mu\,, e_\mu{}^a$ and $m_\mu$ follow
from the known proper transformation rules by omitting the transport term, see eq.~\eqref{rel}:
\begin{align}\label{NCbosbackground}
\begin{split}
 \delta_{{\rm bg}} \tau_\mu &= \partial_\mu\xi^\rho\,\tau_\rho \,,  \\[,1truecm]
 \delta_{{\rm bg}} e_\mu{}^a &= \partial_\mu\xi^\rho\,e_\rho{}^a +\lambda^a{}_b\,e_\mu{}^b
                     +\lambda^a\,\tau_\mu\,, \\[,1truecm]
 \delta_{{\rm bg}} m_\mu &= \partial_\mu\xi^\rho\,m_\rho +\partial_\mu\sigma +\lambda_ae_\mu{}^a \,.
\end{split}
 \end{align}
Here, $\lambda^a{}_b\,, \lambda^a$ and $\sigma$ are the parameters of a local spatial rotation, boost transformation and
central charge transformation, respectively.

The proper transformation rules of the background fields $\tau_\mu\,, e_\mu{}^a$ and $m_\mu$ can be obtained by
gauging the Bargmann algebra, see e.g.~\cite{Andringa:2010it}. Since the NC background is the most general
background one must be able to obtain the transformations of the Curved Galilean and flat backgrounds discussed in the two previous
subsections by gauge-fixing some of the general coordinate transformations. This is discussed in detail in
\cite{Andringa:2013mma}. For the convenience of the reader, we list in table \ref{table:2}
the gauge-fixing conditions that need to be imposed on the NC background fields, and the compensating gauge
transformations that come along with it, that bring us to the Curved Galilean background in terms of the Newton
potential $\Phi$.

\begin{table}[ht]
\begin{center}
{\small
\begin{tabular}{|c|c|}
\hline \rule[-1mm]{0mm}{6mm}
gauge condition & compensating transformations \\[.1truecm]
\hline \rule[-1mm]{0mm}{6mm}

$\tau_\mu(x^\nu)=\delta_\mu{}^\emptyset$ & $\xi^\emptyset(x^\nu)=\xi^\emptyset$ \\[.1truecm]

$\omega_\mu{}^{ab}=0$ & $\lambda^{ab}(x^\nu)=\lambda^{ab}$ \\[.1truecm]

$e_i{}^a=\delta_i{}^a$ & $\xi^i(x^\nu)=\xi^i(t)-\lambda^i{}_jx^j$ \\[.1truecm]

$\tau_i(x^\nu)+m_i(x^\nu)=\partial_im(x^\nu)$ & \\[.1truecm]

$m(x^\nu)=0$ & $\sigma(x^\nu)=\sigma(t)+\partial_t \xi^i(t)\,x^i$ \\[.1truecm]

$\tau_i(x^\nu)=0$ & $\lambda^i(x^\nu)= -\partial_t \xi^i(t)$ \\[.1truecm]

\hline

\hline
$m_\emptyset(x^\nu)=\Phi(x^\nu)$ & $\omega_\emptyset{}^a=-\partial_a\Phi(x^\nu)$ \\[.1truecm]

\hline

\end{tabular}
}
\end{center}
 \caption{\sl This table indicates the gauge-fixing conditions, chronologically ordered from top to bottom, and the
              corresponding compensating transformations, that lead from the NC particle, see subsection \ref{subsec:NC},
              to the Curved Galilean particle, see subsection \ref{subsec:new}. Note that $\tau_i(x^\nu)\equiv e_i{}^0(x^\nu)$.
              The restriction $\tau_i(x^\nu)+m_i(x^\nu)=\partial_im(x^\nu)$ follows from the gauge conditions made at that point.
              }\label{table:2}
\end{table}

\subsection{Adding a Cosmological Constant (bosonic)}\label{subsec:CC}

We now extend the analysis and consider  particles in a  background with a  cosmological constant $\Lambda$.
We only consider the case of a negative cosmological constant.
One way to obtain  the action and transformation rules is by using the method of non-linear realizations
\cite{Coleman:1969sm,Callan:1969sn} applied to the non-relativistic contraction of the (A)dS algebra which is the
so-called Newton--Hooke algebra. In the supersymmetric case this is done in appendix \ref{sec:non-linear_realization}.
The corresponding Newton--Hooke superalgebra is given in appendix \ref{app:A}.
As we will see below, in the bosonic case an easier way of getting a cosmological extension
is to take the non-relativistic limit, as described e.g.~in \cite{Andringa:2012uz}, of the relativistic particle
action in an (A)dS background. Below we will discuss the cosmological extension of the  Galilean,
Curved Galilean and NC particle, one after the other. Following the nomenclature given in table \ref{table:1},
this will lead to the NH, Curved NH and NC NH particle, respectively.

\subsubsection*{The Newton--Hooke Particle}

We consider a negative cosmological constant $\Lambda<0$ with the AdS radius given by
$R^2=-1/\Lambda$ \footnote{In the following we will express everything in terms of the  radius $R$.}.
In global coordinates the metric of an AdS spacetime is given by
\begin{align}
 ds^2=-f(r)\,dt^2+\frac{1}{f(r)}\,dr^2+r^2d\varphi^2 \,, \qquad \qquad f(r)=1-\Lambda r^2 \,.
\end{align}
By taking the non-relativistic limit of a relativistic particle in such an AdS background we arrive at the action
\begin{align}\label{nh-bosonicparticle}
 S= \int\, d\tau\, \frac{m}{2}\, \left[\frac{\dot x^i\dot x^i}{\dot t} -\frac{\dot t \,x^ix^i}{R^2}\right]\,,
\end{align}
once a total derivative term is eliminated.
Physically, this system is equivalent to the non-relativistic harmonic oscillator.
This action is invariant under a deformation of the  Galilei transformations \eqref{Galtrans}. In particular, the
transformations under boosts and spatial translations are given by
\begin{align}\label{nh-trans}
\delta x^i = - v^iR\sin\frac{t}{R} -a^i\cos\frac{t}{R} \,.
\end{align}
The other Galilei transformations are not deformed.
The symmetry algebra is the NH algebra, which consists of the Galilei algebra \eqref{Galalg} plus the
additional commutator
\begin{align}\label{nh-alg}
 \big[P_a,H \big] = \frac{1}{R^2}\,G_a \,.
\end{align}
In the limit $R\to\infty$ the action, transformation rules and the algebra reduce to those of the Galilean particle.

\subsubsection*{The Curved Newton--Hooke Particle}

We can partially gauge the symmetries of the NH particle in the same way as we did in the case of  the Galilean
particle, see subsection \ref{subsec:new}, by introducing a gravitational potential $\phi$.
The resulting action is given by
\begin{align}\label{bosgalNHparticle}
  S &= \int\,d\tau\,\frac{m}{2}\,\left[\frac{\dot x^i\dot x^i}{\dot t} -\dot t\,\frac{x^ix^i}{R^2} -2\,\dot t\,\phi\right] \,.
\end{align}
It is invariant under the acceleration-extended symmetries \eqref{Newtrans} but the transformation of the
background field changes:
\begin{align}\label{phitransNH}
 \delta_{{\rm bg}} \phi = \frac{1}{\dot t}\frac{d}{d\tau}\left(\frac{\dot\xi^i}{\dot t}\right)x^i +\frac{\xi^i(t)\,x^i}{R^2} +\sigma(t)\,.
\end{align}
The explicit $1/R$ terms that appear in the action \eqref{bosgalNHparticle} and the transformation rule
\eqref{phitransNH} are due to our choice of the gravitational potential $\phi$. Namely, we demand that the equation
of motion of the potential is $\partial^i\partial_i\phi=0$.\footnote{\label{footnote}
Note that we could simplify the formulas \eqref{bosgalNHparticle} and \eqref{phitransNH} by defining a new potential
\begin{align}\label{dontdothis}
\tilde \Phi=\phi+\frac{x^ix^i}{2R^2} \,.
\end{align}
This would remove all $1/R$ terms in \eqref{bosgalNHparticle} and \eqref{phitransNH} but it would also alter the equation
of motion. By our definition $\tilde\Phi$ would not be a gravitational potential. Nevertheless, for technical reasons
it is useful to define such a shifted potential, especially in the supersymmetric case.}

Setting the gravitational potential $\phi$ to zero leads us back to the NH particle. The differential equation that
follows from eq.~\eqref{phitransNH} then fixes the time-dependence of $\xi^i(t)$ to be the one given in eq.~\eqref{nh-trans}.

\subsubsection*{The Newton--Cartan Newton--Hooke Particle}

A cosmological extension of the NC particle discussed in subsection \ref{subsec:NC} can  be obtained along the same
lines as we discussed for the Curved NH particle above. The following action is obtained:
\begin{align}\label{NHNCaction}
S = \int\, d\tau\,
\frac{m}{2}\,\left[\frac{\dot x^\mu e_\mu{}^a\,\dot x^\nu e_{\nu\,a}}{\dot x^\rho\tau_\rho} -2\,m_\mu\dot x^\mu
              -\dot x^\rho \tau_\rho \,\frac{x^\mu e_\mu{}^a x^\nu e_{\nu a}}{R^2}\right]\,.
\end{align}
This action is invariant under the gauged Galilei symmetries (which are equivalent to the gauged NH symmetries).
The transformations of the background fields $\tau_\mu$ and $e_\mu{}^a$ are given by eq.~\eqref{NCbosbackground}, while
the transformation of the central charge gauge field $m_\mu$  reads
\begin{align}\begin{split}\label{NHmutrafo}
 \delta_{{\rm bg}} m_\mu &= \partial_\mu\sigma +\lambda^ae_{\mu a}
          -\tau_\mu\,\frac{x^\rho e_{\rho a} x^\nu}{R^2}\,\lambda^a\tau_\rho  \\
       &\quad + m_\rho\partial_\mu\xi^\rho    +\tau_\mu\,\frac{x^\nu e_\nu{}^a}{R^2}\,\xi^\rho e_\rho{}^a
          -\tau_\mu\,\frac{x^\nu e_\nu{}^a}{R^2}\,x^\rho e_\alpha{}^a\partial_\rho\xi^\alpha \,.
\end{split}\end{align}
By imposing the gauge-fixing conditions enlisted in table \ref{table:2} one obtains the action and symmetries of the Curved NH particle, see eqs.~\eqref{bosgalNHparticle} and \eqref{phitransNH}.

\bigskip\noindent
This finishes our discussion of the bosonic  particle in different  backgrounds. In the remaining part of this work
we will discuss the supersymmetric extension of these different particles and backgrounds.

%%%%%%%%%%%%%%%%%%%%%%%%%%%%%%%%%%%%%%%%%%%%%%%%%%%%%%%%%%%%%%%%%%%%%%%%%%%%%%%%%%%%

\section{The Non-relativistic Superparticle}\label{sec:superparticle}

In the same way that the non-relativistic bosonic particle is based upon the Galilei algebra, or its centrally extended
version, the Bargmann algebra, the action and transformation rules of the non-relativistic 3D $\mathcal{N}=2$ superparticle
are based upon the supersymmetric extension of the Galilei or Bargmann algebra. It turns out  that we need {\sl two} supersymmetries
since one of the supersymmetries is, like the time translations in the bosonic case, a St\"uckelberg symmetry.

The 3D $\mathcal{N}=2$ Galilei superalgebra is given by the bosonic commutation relations \eqref{Galalg} plus the additional
relations \cite{Bergman:1995zr,Bergman:1996bx}
\begin{align}\begin{split}\label{Bargmannalgebra}
 \big[J_{ab},Q^\pm \big] &= -\frac12\,\gamma_{ab} \,Q^\pm \,, \hskip2.3truecm \big[G_a,Q^+ \big] = -\frac12\,\gamma_{a0}\,Q^- \,,\\[.1truecm]
 \big\{Q^+_\alpha,Q^+_\beta \big\} &= 2\,[\gamma^0C^{-1}]_{\alpha\beta}\,H \,,
      \hskip1.5truecm \big\{Q^+_\alpha,Q^-_\beta \big\} =[\gamma^aC^{-1}]_{\alpha\beta}\,P_a \,,
\end{split}\end{align}
where the Majorana spinors $Q_\alpha^\pm$ are the generators corresponding to the two supersymmetries.\,\footnote{We use
a Majorana representation in which the charge conjugation matrix is given by $C=i\gamma^0$ and all
$\gamma$-matrices are real, i.e.~$\gamma^\mu=(i\sigma_2,\sigma_1,\sigma_3)$.} In the case of the 3D $\mathcal{N}=2$
Bargmann superalgebra there is a central extension
given by eq.~\eqref{pqzcommutator} and by the following equation:
\begin{align}
 \big\{Q^-_\alpha,Q^-_\beta \big\} = 2\,[\gamma^0C^{-1}]_{\alpha\beta}\,Z \,.
\end{align}

The 3D $\mathcal{N}=2$ supergravity background has been obtained in \cite{Andringa:2013mma} by gauging the algebra above. This
result allows us to discuss the non-relativistic superparticle for $\Lambda=0$, i.e.~the supersymmetric version of the
second column of table \ref{table:1}. In the three subsections below we will discuss the Galilean, the Curved Galilean
and the NC superparticle, respectively.

In the relativistic case the superparticle has an additional so-called $\kappa$-symmetry. In the non-relativistic analog
this becomes a St\"uckelberg symmetry \cite{Gomis:2004pw}. For the purpose of this section we will fix
$\kappa$-symmetry to avoid unnecessary
cluttering of our formulas. Some remarks about restoring $\kappa$-symmetry are separately given in section \ref{sec:kappa}.

\subsection{The Galilean Superparticle}\label{sec:supergal}

The Galilean superparticle was already discussed in \cite{Gomis:2004pw}.
In terms of the bosonic and fermionic embedding coordinates $\{t,x^i,\theta_-\}$ the action is given by
\begin{align}\label{NRnokappa}
  S= \int d\tau\,  \frac{m}{2}\,\left[ \frac{\dot x^i\dot x^i}{\dot t} -\bar\theta_-\gamma^0\dot\theta_-\right] \,.
\end{align}
This action is invariant under the following Galilei  symmetries:
\begin{align}\begin{split}\label{bosgaugefixedNRtrafo}
 \delta t = -\zeta \,, \hskip1truecm \delta x^i = \lambda^i{}_j x^j -v^i t -a^i \,. \hskip1truecm
 \delta \theta_- = \frac14\lambda^{ab}\gamma_{ab}\theta_- \,.
\end{split}\end{align}
The same action is invariant under two supersymmetries with constant parameters $\epsilon_+$ and $\epsilon_-$:
 \begin{align}\begin{split}\label{fermgaugefixedNRtrafo}
 \delta t &= 0 \,, \hskip1.5truecm
 \delta x^i = -\frac12\bar\epsilon_+\gamma^i\theta_-\,, \hskip 1.5truecm
 \delta \theta_- = \epsilon_- -\frac{\dot x^i}{2\dot t}\gamma_{0i}\epsilon_+ \,.
\end{split}\end{align}
One may verify that the above set of transformation rules \eqref{bosgaugefixedNRtrafo} and \eqref{fermgaugefixedNRtrafo}
closes off-shell.
Note that the transformation with parameter $\epsilon_+$ is realized linearly. Instead, the one with parameter
$\epsilon_-$ is realized non-linearly, i.e.~it is a broken symmetry. This implies that the superparticle
corresponds to a  $1/2$ BPS state.

\subsection{The Curved Galilean Superparticle}\label{sec:supernew}

We will now extend the Galilean superparticle to a Curved Galilean superparticle thereby replacing the flat background by a
 Galilean supergravity background. This corresponds to extending the bosonic particle in a  Galilean
gravity background, discussed in section \ref{subsec:new}, to the supersymmetric case.

Our starting point is the superparticle action in a flat background, see eq.~\eqref{NRnokappa}.
We will now  partially gauge the transformations
\eqref{bosgaugefixedNRtrafo} and \eqref{fermgaugefixedNRtrafo} to allow for arbitrary time-dependent boosts,
with parameters $\xi^i(t)$, and arbitrary supersymmetry transformations, with parameters $\epsilon_-(t)$. The complete
bosonic and fermionic transformation rules now read
\begin{align}\begin{split}\label{bostdeptrafo}
 \delta t = -\zeta \,, \hskip1truecm  \delta x^i = \lambda^i{}_j x^j -\xi^i(t) \,, \hskip1truecm
 \delta \theta_- = \frac14\lambda^{ab}\gamma_{ab}\theta_- \,,
\end{split}\end{align}
and
\begin{align}\begin{split}\label{fermtdeptrafo}
 \delta t &= 0 \,, \hskip 1.5truecm
 \delta x^i = -\frac12\bar\epsilon_+\gamma^i\theta_- \,, \hskip 1.5truecm
 \delta \theta_- = \epsilon_-(t) -\frac{\dot x^i}{2\dot t}\gamma_{0i}\epsilon_+ \,,
\end{split}\end{align}
respectively.

The Galilean supergravity multiplet, that we will use to perform the partial gauging of the transformations
\eqref{bosgaugefixedNRtrafo} and \eqref{fermgaugefixedNRtrafo}, was introduced in \cite{Andringa:2013mma}.
Alongside the Newton potential $\Phi(x)$, it also contains a fermionic background field $\Psi(x)$. The
equations of motion for these two background fields are:
\begin{align}\label{eomphipsi}
 \partial^i\partial_i\Phi=0\,, \hskip1.5cm \gamma^i\partial_i\Psi=0 \,.
\end{align}
There is a slight subtlety regarding this Galilean supergravity multiplet, stemming from the fact that $\Psi(x)$
is the superpartner of the Newton force $\Phi_i \equiv \partial_i \Phi(x)$ and not of the Newton potential itself.
The transformation rules of the Newton force are, however, compatible with the integrability condition
$\partial_{[i} \Phi_{j]} = 0$, so that they can be integrated to transformation rules of the Newton potential
$\Phi(x)$. This is done via the introduction of a fermionic prepotential $\chi(x)$, that will be called the
`Newtino potential', defined via
\begin{align}\label{psichirel}
\partial_i\,\chi =\gamma_i\Psi\ \ \ (i=1,2) \,,\hskip 1.5truecm \gamma^1\partial_1\chi = \gamma^2\partial_2\chi\,,
\end{align}
where the second equation represents a constraint obeyed by $\chi(x)$, as a consequence of its definition.
This constraint can be interpreted (upon choosing a specific basis for the $\gamma$-matrices) as the Cauchy--Riemann
equations, expressing holomorphicity of $\chi_1 + \rmi\, \chi_2$, where $\chi_{1,2}$ are the components of
$\chi$. Since the Newton potential $\Phi$ obeys the Laplace equation in two spatial dimensions, it can also be
seen as the real part of a holomorphic function $\Phi + \rmi \, \Xi$. The imaginary part  $\Xi(x)$ of this
function was called the `dual Newton potential' in \cite{Andringa:2013mma} and is related to $\Phi(x)$ via
the Cauchy--Riemann equations for $\Phi + \rmi \, \Xi$:
\begin{align}\label{phixirel}
 \partial_i\Phi &= \varepsilon_{ij}\partial^j\Xi \,, \hskip1.5cm  \partial_i\Xi = -\varepsilon_{ij}\partial^j\Phi \,.
\end{align}
The dual Newton potential was introduced in \cite{Andringa:2013mma} in order to write down the supersymmetry
transformation rule for $\chi$. Since this is a transformation rule for both real and imaginary parts of
$\chi_1 + \rmi\, \chi_2$, it is natural to expect that it involves also both real and imaginary parts of
$\Phi + \rmi \, \Xi$ and this is indeed the case.

We find that the action of the Curved Galilean superparticle in terms of the Galilean supergravity background
fields $\Phi$ and $\Psi$ is given by
\begin{align}\label{galileanparticle}
  S &= \int d\tau\,\frac{m}{2}\,\left[\frac{\dot x^i\dot x^i}{\dot t} -\bar\theta_-\gamma^0\dot\theta_- -2\dot t\, \Phi
         +2\dot t\, \bar\theta_-\gamma^0\Psi \right]\,.
\end{align}
One may verify that the action \eqref{galileanparticle}
is invariant under the transformations \eqref{bostdeptrafo} and \eqref{fermtdeptrafo}
provided that  the background fields transform under the bosonic symmetries as
\begin{align}\begin{split}\label{phipsibostrans}
 \delta_{\rm bg} \Phi &= \frac{1}{\dot t}\frac{d}{d\tau}\left(\frac{\dot\xi^i}{\dot t}\right)x^i +\sigma(t)\,, \hskip1.5cm
 \delta_{\rm bg} \Psi = \frac14\lambda^{ab}\gamma_{ab}\Psi \,,
\end{split}\end{align}
and under the fermionic symmetries as
\begin{align}
 \delta_{\rm bg} \Phi &= \bar\epsilon_-\gamma^0\Psi +\frac12\,\bar\epsilon_+\partial_t\chi
                 -\frac12\,\bar\epsilon_+\gamma^i\theta_-\,\partial_i\Phi \,, \label{Phitrafo}  \\
 \delta_{\rm bg} \Psi &= \frac{1}{\dot t}\,\dot\epsilon_- -\frac12\,\partial_i\Phi\gamma_{i0}\epsilon_+
                 -\frac12\,\bar\epsilon_+\gamma^i\theta_-\,\partial_i\Psi \,. \label{chitrafo}
\end{align}

The only invariance that is non-trivial to show is the one under the linear $\epsilon_+$-trans\-for\-mations.
Varying the action \eqref{galileanparticle} under $\epsilon_+$-transformations one is left with the following terms:
\begin{align}
 \delta_+S= \int d\tau\,\frac{m}{2} \Big[-\bar\epsilon_+\,\dot t\,\partial_t\chi  -\bar\epsilon_+\,\dot x^i\,\partial_i \chi
                    -\frac{\dot t}{2}\,\bar\epsilon_+\gamma^k\theta_-\,\bar\theta_-\gamma^{0i}\partial_i\partial_k\chi \Big] \,.
\end{align}
The first two terms combine into a total $\tau$-derivative, since $\chi$ is a function of $x^i$ and $t$ and therefore
\begin{equation}
\frac{d}{d\tau} \,\chi=\big(\dot t\, \partial_t+\dot x^i\,\partial_i\big) \chi\,.
\end{equation}
The second term vanishes upon using the equation of motion for the background field $\chi$.

To calculate the  commutator algebra it is important to keep in mind that the background fields  do not transform
as fundamental fields but, instead, according to background fields, see eq.~\eqref{rel}.
This explains the `wrong' sign transport term in the $\epsilon_+$-transformation and the absence of transport
terms for all other symmetries. It also has the consequence that  partial derivatives
do not commute with background variations:
\begin{align}
 \big[\delta_{\rm bg} ,\partial_\mu \big]=-(\partial_\mu \delta x^\nu)\partial_\nu \,.
\end{align}
Another subtlety when calculating the commutator algebra is related to the fact that the parameters $\xi^i$ and $\epsilon_-$
are functions of the time
$t$ but that $t$ itself is a scalar function $t(\tau)$ of the world-line parameter $\tau$. This implies that when
we calculate commutators we have to vary the $t$ inside the parameters. Keeping the above subtleties in mind we find that
the commutation relations close off-shell on the embedding coordinates and the background fields.

Imposing the gauge-fixing conditions
\begin{align}
 \Phi=1 \,,\hskip1.5cm \chi=0 \,,
\end{align}
we recover the Galilean superparticle with the  flat spacetime transformation rules \eqref{bosgaugefixedNRtrafo}
and \eqref{fermgaugefixedNRtrafo}. Imposing the additional gauge-fixing condition
\begin{align}\label{rhofix}
t=\tau
\end{align}
we find agreement with the algebra obtained in \cite{Andringa:2013mma}.

%%%%%%%%%%%%%%%%%%%%%%%%%%%%%%%%%%%%%%%%%%%%%%%%%%%%%%%%%%%%%%%%%%%%%%%%%%%%%%%%

\subsection{The Newton--Cartan Superparticle}\label{sec:superNC}

We wish to extend the result of the previous subsection to arbitrary frames corresponding to
a superparticle in a Newton--Cartan supergravity background. Due to the complexity of the calculations we only give the
result up to quartic fermions in the action. We find that using this approximation the action is given by
\begin{align}\begin{split}\label{NCparticle}
 S = \int d\tau\,\frac{m}{2}\,\left[\frac{\dot x^\mu e_\mu{}^a\,\dot x^\nu e_{\nu\,a}}{\dot x^\rho\tau_\rho} -2m_\mu\dot x^\mu
                     -\bar\theta_-\gamma^0D_\tau\theta_- +2\,\bar\theta_-\gamma^0 \psi_{\mu-}\dot x^\mu
               -\frac{\dot x^\mu e_{\mu a}}{\dot x^\rho\tau_\rho}\,\bar\theta_-\gamma^a\psi_{\nu+}\dot x^\nu  \right] \,,
\end{split}\end{align}
where the Lorentz-covariant derivative $D_\tau$ is defined as
\begin{align}
 D_\tau \theta_-=\dot\theta_--\frac{1}{4}\,\dot x^\mu\omega_\mu{}^{ab}\gamma_{ab}\theta_- \,.
\end{align}
To lowest order in the fermions the action \eqref{NCparticle} is invariant under the following  bosonic
and fermionic symmetries of the embedding coordinates:
\begin{align}\begin{split}
 \delta x^\mu &= -\xi^\mu(x^\alpha) \,, \hskip1.5truecm
 \delta \theta_- = \frac14\,\lambda^{ab}(x^\alpha)\gamma_{ab}\,\theta_- \,,
\end{split}\end{align}
and
\begin{align}\begin{split}\label{coordQ}
 \delta x^\mu &= -\frac12\,\bar\epsilon_+(x^\alpha)\gamma^a\theta_-\,e^\mu{}_a \,, \hskip1.5truecm
 \delta \theta_- = \epsilon_-(x^\alpha)
        -\frac{\dot x^\mu e_\mu{}^a}{2\dot x^\rho\tau_\rho}\,\gamma_{0a}\,\epsilon_+(x^\alpha) \,.
\end{split}\end{align}
In the following we refrain from explicitly denoting the local $x^\mu$-dependence of the para\-meters.

The transformation rules of the background fields follow from the supergravity result given in \cite{Andringa:2013mma}
and application of the identity \eqref{rel}. We find that the bosonic transformation rules are given by
\begin{align}\begin{split}\label{NCbos}
 \delta_{\rm pr} \tau_\mu &= 0 \,, \hskip4.4truecm       \delta_{\rm pr} m_\mu = \partial_\mu\sigma +\lambda_ae_\mu{}^a \,, \\[.1truecm]
 \delta_{\rm pr} e_\mu{}^a &= \lambda^a{}_b\,e_\mu{}^b   +\lambda^a\,\tau_\mu\,, \hskip2truecm
          \delta_{\rm pr} \psi_{\mu+} = \frac14\lambda^{ab}\gamma_{ab}\psi_{\mu+} \,, \\
 \delta_{\rm pr} \omega_\mu{}^{ab} &=   \partial_\mu\lambda^{ab} \,, \hskip3.5truecm
          \delta_{\rm pr} \psi_{\mu-} = \frac14\lambda^{ab}\gamma_{ab}\psi_{\mu-}
                     -\frac12\lambda^a\gamma_{a0}\psi_{\mu+} \,,\\[.1truecm]
 \delta_{\rm pr} \omega_\mu{}^a &=
       \partial_\mu\lambda^a-\lambda_b\omega_\mu{}^{ab} +\lambda^{ab}\omega_{\mu b}\,.
\end{split}\end{align}
To keep the formulas simple we have given here as well as below only the proper transformation rules. The background
transformations are obtained by supplementing each of these rules with an additional transformation under general
coordinate transformations, see eq.~\eqref{rel}. For the fermionic transformations we find the following expressions:
\begin{align}\begin{split}\label{NCferm}
 \delta_{\rm pr} \tau_\mu &= \frac12\,\bar\epsilon_+\gamma^0\psi_{\mu+}\,, \hskip3.95truecm
               \delta_{\rm pr} m_\mu = \bar\epsilon_-\,\gamma^0\psi_{\mu-} \,, \\
 \delta_{\rm pr} e_\mu{}^a &= \frac12\,\bar\epsilon_+\gamma^a\psi_{\mu-} +\frac12\,\bar\epsilon_-\gamma^a\psi_{\mu+}\,, \hskip1.52truecm
               \delta_{\rm pr} \psi_{\mu+} = D_\mu\epsilon_+  \,, \\
 \delta_{\rm pr} \omega_\mu{}^{ab} &= 0\,, \hskip5.45truecm
                \delta_{\rm pr} \psi_{\mu-} = D_\mu\epsilon_-  +\frac12\,\omega_\mu{}^a\gamma_{a0}\epsilon_+ \,.
\end{split}\end{align}
The variation of $\omega_\mu{}^{ab}$ is only zero on-shell, i.e.~upon using the equations of motion of
the background fields. The explicit form of these equations of motion are given in \cite{Andringa:2013mma}. They
also follow by taking the $R\to \infty$ limit of the formulas in appendix \ref{app:NCNHparticle}.
In the same manner we can write the variation of $\omega_\mu{}^a$ as
\begin{align}\label{omegaQ}
 \delta_{\rm pr} \omega_\mu{}^a = \frac12\,\bar\epsilon_-\gamma^0\hat\psi_\mu{}^a{}_-
        +\frac12\,\tau_\mu\,\bar\epsilon_-\gamma^0\hat\psi_0{}^a{}_- +\frac14\,e_\mu{}^b\,\bar\epsilon_+\gamma^b\hat\psi^a{}_{0-}
        +\frac14\,\bar\epsilon_+\gamma^a\hat\psi_{\mu0-} \,,
\end{align}
where $\hat\psi_{\mu\nu-}$ is the covariant curvature of $\psi_{\mu-}$, see  \cite{Andringa:2013mma} or
appendix \ref{app:NCNHparticle} with $R\to\infty$. One may check that, to lowest order in fermions, the action
\eqref{NCparticle} is invariant under the transformations \eqref{coordQ}, \eqref{NCferm} and \eqref{omegaQ},
upon use of the equations of motion of the background fields.

As a consistency check we  have verified that by imposing the gauge-fixing conditions\,\footnote{For the bosonic
case these conditions are given in table \ref{table:2}.} of \cite{Andringa:2013mma} the action and transformation
rules of the NC superparticle reduce to those of the Curved Galilean superparticle.

%%%%%%%%%%%%%%%%%%%%%%%%%%%%%%%%%%%%%%%%%%%%%%%%%%%%%%%%%%%%%%%%%%%%%%%%

\section{Adding a Cosmological Constant}\label{sec:superCC}

In this section we are going to describe the superparticle in the presence of a cosmological constant.
Like in the bosonic case we can derive the action by taking the non-relativistic limit of a superparticle in AdS
space, see e.g.~\cite{Gomis:2005pg} for an example in 10 dimensions.

In the presence of a cosmological constant the relativistic AdS superalgebra in three dimensions is not unique. Instead, in the
case of $\mathcal{N}$ supersymmetries, one always finds $\mathcal{N}$
different versions, often referred to as $(p,q)$  AdS superalgebras \cite{Achucarro:1987vz}.
In appendix \ref{app:A} we give both the $(1,1)$ and $(2,0)$ $\mathcal{N}=2$ AdS superalgebras. As  explained in appendix \ref{app:A},
the Newton--Hooke superalgebra that we will use below is obtained by contracting the  $\mathcal{N}=(2,0)$ AdS superalgebra.

\subsection{The Newton--Hooke Superparticle}\label{sec:superNH}

We find that the action of the  NH superparticle takes the form
\begin{align}\label{snh-particleaction}
 S=\int d\tau \,\frac{m}{2}\,\left[\frac{\dot x^i\dot x^i}{\dot t} -\bar\theta_-\gamma^0\dot \theta_-
                   -\frac{\dot t\,x^ix^i}{R^2} +\frac{3\dot t}{2R}\,\bar\theta_-\theta_- \right]\,.
\end{align}
A realization of the Newton--Hooke superalgebra, whose explicit form is given in  eq.~\eqref{n20-nh-}, on the
embedding coordinates is given by the following bosonic transformation rules
\begin{align}\begin{split}\label{snh-bostrans}
 \delta t=-\zeta \,, \qquad \delta x^i = \lambda^i{}_kx^k -\xi^i(t) \,, \qquad
 \delta \theta_-=\frac14\lambda^{ab}\gamma_{ab}\theta_- \,,
\end{split}\end{align}
supplemented with the following fermionic transformations:
\begin{align}\begin{split}\label{snh-fermtrans1}
 \delta t=0 \,, \qquad \delta x^i = -\frac12\,\bar\epsilon_+(t)\gamma^i\theta_- \,, \qquad
 \delta \theta_-=\epsilon_-(t) -\frac{\dot x^i}{2\dot t}\,\gamma_{0i}\,\epsilon_+(t) +\frac{x^i}{2R}\,\gamma_i\,\epsilon_+(t) \,.
\end{split}\end{align}
Here, the time-dependence of the parameters $\xi^i(t)$ and $\epsilon_\pm(t)$ is given by
\begin{align}\begin{split}\label{snh-timedep}
 \xi^i(t) &= v^iR\sin\frac{t}{R} +a^i\cos\frac{t}{R}\,, \\
 \epsilon_-(t) &={\rm exp}\Big(\frac{3t}{2R}\,\gamma_0\Big)\,\epsilon_- \,, \qquad\qquad
 \epsilon_+(t)={\rm exp}\Big(-\frac{t}{2R}\,\gamma_0\Big)\,\epsilon_+ \,.
\end{split}\end{align}
Like in the $\Lambda=0$ case, the $\epsilon_+$-transformation  is realized linearly while the $\epsilon_-$-trans\-for\-mation,
corresponding to a broken supersymmetry, is not. The superparticle  thus corresponds to a $1/2$ BPS state.
We have verified that the transformations \eqref{snh-bostrans} and \eqref{snh-fermtrans1} leave the NH superparticle action
\eqref{snh-particleaction} invariant.

In the formulation we are using both supersymmetries are time-dependent but $\theta_-$ is invariant under constant time shifts.
Alternatively, one can absorb the time-dependence of either $\epsilon_+$ or $\epsilon_-$, but not both at the same time, in
a redefinition of $\theta_-$. Such a redefinition would
introduce a non-vanishing transformation $\delta_\zeta\tilde\theta_-$ and shift the $\bar\theta_-\theta_-$ term in the action.
We prefer to keep the time-dependent description of eqs.~\eqref{snh-particleaction}--\eqref{snh-timedep}.

\subsection{The Curved Newton--Hooke Superparticle}\label{subsec:superNHgal}

The Curved Newton--Hooke superparticle is obtained in analogy to the Curved Galilean superparticle, see subsection
\ref{sec:supernew}. This means that we gauge the spatial translations and one of the supersymmetries, the broken one,
such that their parameters become  arbitrary time-dependent
functions. This introduces the Newton potential $\phi$ and its supersymmetric partner $\psi$. However, based upon
our experience in the bosonic case, we expect that either the transformation rules of the NH background fields
$\phi$ and $\psi$ differ from the Galilean background fields $\Phi$ and $\Psi$, or their equations of motion change,
see the comment in footnote \ref{footnote}. Either way, we cannot make use of the results of \cite{Andringa:2013mma}.\footnote{
The equations of motion are needed to close the supersymmetry algebra.}
Instead, we should first derive the transformations
rules of Newton--Hooke supergravity as a $1/R$ modification of Galilean supergravity.
Once we have obtained the NH supergravity transformation rules we can use eq.~\eqref{rel}
to get the transformation rules of the NH background fields.

% \vskip .4truecm
% \centerline{\bf Newton--Hooke supergravity}
% \vskip .2truecm
\subsubsection*{Newton--Hooke supergravity}

We find that the following bosonic and fermionic transformation rules provide a realization of the acceleration-extended NH algebra:
\begin{align}\begin{split}\label{NHphipsibostrans}
 \delta_{\rm pr} \phi &= \zeta\partial_t\phi + \xi^i(t)\partial_i\phi +\partial_t\partial_t\xi^i(t)\,x^i +\frac{1}{R^2}\,\xi^i(t)\,x^i
                -\lambda^i{}_jx^j\partial_i\phi +\sigma(t)\,, \\
 \delta_{\rm pr} \psi &= \zeta\partial_t\psi + \xi^i(t)\partial_i\psi +\frac14\lambda^{ab}\gamma_{ab}\psi
                -\lambda^i{}_jx^j\partial_i\psi \,,
\end{split}\end{align}
and
\begin{align}
 \delta_{\rm pr} \phi &= \bar\epsilon_-(t)\gamma^0\psi +\frac12\,\partial_t\big(\bar\epsilon_{+,t}\,\chi\big)
                -\frac{x^i}{2R}\,\bar\epsilon_{+,t}\gamma_{0i}\psi \,, \label{NHPhitrafo}  \\
 \delta_{\rm pr} \psi &= \partial_t\epsilon_-(t) -\frac{3}{2R}\,\gamma_0\epsilon_-(t)
                -\frac12\,\partial_i\phi\,\gamma_{i0}\epsilon_{+,t} \,. \label{NHPsitrafo}
%  \Big(\delta \chi &= x^i\gamma_i\,\big(\partial_t\epsilon_-(t)-\frac{3}{2R}\,\gamma_0\epsilon_-(t)\big)
%                 +\frac12\big(\Xi-\Phi\gamma_0\big)\,\epsilon_{+,t} \Big)\,, \label{NHchitrafo}
\end{align}
These transformation rules constitute the NH supergravity extension of the Galilean supergravity result given in
\cite{Andringa:2013mma}.
The subscript $t$ on $\epsilon_{+,t}$ indicates that it is a function of time. However, unlike
$\xi^i(t)$ or $\epsilon_-(t)$, the time-dependence of $\epsilon_{+,t}$ is a very specific one, namely
the one given in eq.~\eqref{snh-timedep}. The background fields $\phi$ and $\psi$ obey the same equations
of motion as in the Galilean case, see eq.~\eqref{eomphipsi}, and are related to their dual potentials $\Xi$ and
$\chi$ by eqs.~\eqref{phixirel} and \eqref{psichirel}, respectively.

%\vskip .4truecm
\subsubsection*{The superparticle action}

Now that we have constructed the Newton--Hooke supergravity theory, it is  easy to construct an action for the
Curved NH superparticle where the NH supergravity fields occur as background fields. We find that this action  is given by
\begin{align}\label{snh-galparticle}
 S=\int d\tau \,\frac{m}{2}\,\left[\frac{\dot x^i\dot x^i}{\dot t} -\bar\theta_-\gamma^0\dot \theta_-
                   -\frac{\dot t\,x^ix^i}{R^2} +\frac{3\dot t}{2R}\,\bar\theta_-\theta_-
                   -2\dot t\, \phi +2\dot t \,\bar\theta_-\gamma^0\psi \right]\,.
\end{align}
The transformation rules of the NH background fields follow from the NH supergravity rules given in
eqs.~\eqref{NHphipsibostrans}--\eqref{NHPsitrafo} after applying the relation \eqref{rel}.

As far as we know, for $\Lambda \ne 0$ there is no redefinition akin to \eqref{dontdothis} that would enable us
to deduce the transformation rules of the Curved Newton--Hooke background fields $\phi$ and $\psi$ from the Curved
Galilean ones given in eqs.~\eqref{Phitrafo} and \eqref{chitrafo},  respectively.

\subsection{The Newton--Cartan Newton--Hooke Superparticle}\label{subsec:superNHNC}

To obtain the NC NH superparticle action we first need to derive the transformation rules of NC NH supergravity, viewed
as a $1/R$ modification of the NC supergravity theory constructed in \cite{Andringa:2013mma}. We find the
following result.

\subsubsection*{Newton--Cartan Newton--Hooke supergravity}

The bosonic transformation rules of the background fields are given by \eqref{NCbos} and \eqref{NHmutrafo}.
With respect to supersymmetry, the only $1/R$ modifications occur in the
transformation rules of $m_\mu$, $\psi_{\mu+}$ and $\psi_{\mu -}$:
\begin{align}
 \delta_{\rm pr} m_\mu &= \bar\epsilon_-\gamma^0\psi_{\mu-}
                -\frac12\,\bar\epsilon_+\gamma^a\theta_-e^\rho{}_a\partial_\rho m_\mu \\
              &\quad  -\frac{(x^\nu e_\nu{}^a)^2}{4R^2}\,\bar\epsilon_+\gamma^0\psi_{\mu+}
                -\tau_\mu\,\frac{x^\nu x^\rho e_\nu{}^a}{2R^2}\,\big(\bar\epsilon_-\gamma^a\psi_{\rho+}
                                                                +\bar\epsilon_+\gamma^a\psi_{\rho-}\big) \,\nonumber\\
 \delta_{\rm pr} \psi_{\mu+} &= D_\mu\epsilon_+ +\frac{1}{2R}\,\tau_\mu\gamma_0\epsilon_+ \,,\label{NHdelpsi+}\\
 \delta_{\rm pr} \psi_{\mu-}&= D_\mu\epsilon_- +\frac12\,\omega_\mu{}^a\gamma_{a0}\epsilon_+
                -\frac{3}{2R}\,\tau_\mu\gamma_0\epsilon_- -\frac{1}{2R}\,e_\mu{}^a\gamma_a\epsilon_+\,.
\label{NHdelpsi-}
\end{align}
The supersymmetry rules of $\tau_\mu,e_\mu{}^a$ are un-deformed and coincide with the ones of NC supergravity,
see \eqref{NCferm}. The above modifications induce the following $1/R$ modifications in the
transformation rules of the dependent fields $\omega_\mu{}^{ab}$ and $\omega_\mu{}^a$:
\begin{align}
 \delta_{\rm pr} \omega_\mu{}^{ab}&=-\frac{\varepsilon_{ab}}{2R}\,\bar\epsilon_+\gamma^0\psi_{\mu+} \,,\\[.2truecm]
\begin{split}\label{NHomegaQ}
 \delta_{\rm pr}\omega_\mu{}^a &= \frac12\,\bar\epsilon_-\gamma^0\hat\psi_\mu{}^a{}_-
        +\frac12\,\tau_\mu\,\bar\epsilon_-\gamma^0\hat\psi_0{}^a{}_- +\frac14\,e_\mu{}^b\,\bar\epsilon_+\gamma^b\hat\psi^a{}_{0-}
        +\frac14\,\bar\epsilon_+\gamma^a\hat\psi_{\mu0-} \\
     &\quad +\frac{1}{2R}\,\bar\epsilon_-\gamma^{a0}\psi_{\mu+} +\frac{1}{2R}\,\bar\epsilon_+\gamma^{a0}\psi_{\mu-} \,.
\end{split}\end{align}

\subsubsection*{The superparticle action}

Having constructed the NC NH supergravity theory, it is straightforward to construct the action for the NC NH
superparticle. We find that the action, up to quartic fermions,  is given by
\begin{align}
 S = \int d\tau\,\frac{m}{2}\,\bigg[&\frac{\dot x^\mu e_\mu{}^a\,\dot x^\nu e_{\nu\,a}}{\dot x^\rho\tau_\rho}
                     -2 m_\mu\dot x^\mu
                     -\bar\theta_-\gamma^0D_\tau\theta_- +2\,\bar\theta_-\gamma^0\psi_{\mu-}\dot x^\mu \nonumber\\
     &\quad          -\frac{\dot x^\mu e_{\mu a}}{\dot x^\rho\tau_\rho}\,\bar\theta_-\gamma^a\psi_{\nu+}\dot x^\nu
     -\dot x^\rho\tau_\rho \,\frac{(x^\mu e_\mu{}^a)^2}{R^2}\,
           +\frac{3}{2R}\,\dot x^\rho\tau_\rho \,\bar\theta_-\theta_- \bigg]  \,.  \label{NHparticle}
\end{align}
One can show that, upon using the equations of motion of the background fields, see appendix \ref{app:NCNHparticle},
this action is indeed invariant up to quartic fermions. This concludes our description of the NC NH superparticle action.

%%%%%%%%%%%%%%%%%%%%%%%%%%%%%%%%%%%%%%%%%%%%%%%%%%%%%%%%%%%%%%%%%%%%%%%%%%%%%%%%%%%

\section{Kappa-symmetry}\label{sec:kappa}

It is well-known that the relativistic superparticle is invariant under an additional infinitely-reducible
fermionic symmetry called $\kappa$-symmetry \cite{deAzcarraga:1982dw,Siegel:1983hh}. In the case of strings and branes,
this $\kappa$-symmetry is needed to obtain the
correct counting of degrees of freedom. It is known from the work of \cite{Gomis:2004pw} that in the non-relativistic
case the $\kappa$-symmetry is just a St\"uckelberg symmetry that acts as a shift on one of the
fermionic coordinates. It can easily  be gauge-fixed upon setting that specific fermionic coordinate equal to zero.
Nevertheless, to obtain an elegant superspace description of the non-relativistic superparticle it might be advantageous
to retain this extra fermionic coordinate since it plays the role  of one of the superspace coordinates.
For this reason we give here the results with $\kappa$-symmetry in  two simple cases,
namely the  Galilean and the  NH superparticle. The action and transformation rules of both superparticles are derived
in appendix \ref{sec:non-linear_realization} by applying the technique of nonlinear realizations.
Below we just give the results thereby focusing  our attention on the $\kappa$-symmetry aspects. For the full details
we refer the reader to appendix \ref{sec:non-linear_realization}.

\subsubsection*{The $\kappa$-symmetric  Galilean superparticle}

The $\kappa$-symmetric version of the flat Galilean superparticle action \eqref{NRnokappa} depends on an additional
fermionic coordinate $\theta_+$. The action is given by
\begin{align}\label{NRparticle}
 S= \int d\tau\, \frac{m}{2}\,\left[ \frac{\pi^i\pi^i}{\pi^0} -\bar\theta_-\gamma^0\dot\theta_-\right] \,,
\end{align}
where the line-elements $\pi^0$ and $\pi^i$ are defined as
\begin{align}\begin{split}\label{lineelem}
 \pi^0 = \dot t  +\frac14\bar\theta_+\gamma^0\dot\theta_+ \,, \qquad\quad
 \pi^i &= \dot x^i  +\frac14\bar\theta_-\gamma^i\dot\theta_+  +\frac14\bar\theta_+\gamma^i\dot\theta_- \,.
\end{split}\end{align}

The $\kappa$-transformations that leave the action \eqref{NRparticle} invariant are given by
\begin{align}\begin{split}\label{kappatrans}
 \delta_\kappa t &= \frac14\bar\kappa\,\gamma^0\theta_+  \,, \qquad
 \delta_\kappa x^i = \frac14\bar\kappa\,\gamma^i\theta_- -\frac{\pi^k}{8\pi^0}\,\bar\kappa\,\gamma_k\gamma^0\gamma^i\theta_+ \,,\\[.2truecm]
 \delta_\kappa \theta_+ &= \kappa \,, \hskip 1.85truecm
 \delta_\kappa \theta_- = -\frac{\pi^i}{2\pi^0}\gamma^0\gamma_i\kappa \,.
\end{split}\end{align}
The commutator algebra of all symmetry transformations including local $\kappa$-transformations and
worldline reparameterizations closes on-shell.

We see from eq.~\eqref{kappatrans} that the $\kappa$-transformation acts as a simple St\"uckelberg shift symmetry
on the embedding coordinate $\theta_+$. Therefore, one could  just fix this symmetry by imposing the
gauge condition
\begin{align}
 \theta_+ = 0 \,,
\end{align}
which leads us back to the formulas of the  Galilean superparticle used in subsection \ref{sec:supergal}.

\subsubsection*{The $\kappa$-symmetric  Newton--Hooke superparticle}

An analogous construction leads to the $\kappa$-symmetric Newton--Hooke superparticle.
We find the following action
\begin{align}\label{kappaNH1}
 S&= \int d\tau\,\frac{m}{2}\,\left[\frac{\pi^i\pi^i}{\pi^0}-\bar\theta_-\gamma^0\dot\theta_-
            -\dot t\,\frac{x^ix^i}{R^2}+\frac{3\dot t}{2R}\,\bar\theta_-\theta_-
            +\frac{\dot t\,x^i}{2R^2}\,\bar\theta_+\gamma^i\theta_-
            -\frac{\dot t}{16R^2}\,\bar\theta_+\theta_+\,\bar\theta_-\theta_- \right] \,,
\end{align}
with the line-elements given by
\begin{align}
 &\pi^0 = \dot t\,\big(1+\frac{1}{8R}\,\bar\theta_+\theta_+\big) +\frac14\,\bar\theta_+\gamma^0\dot\theta_+ \,, \\
 &\pi^i = \big(\dot x^i+\frac14\,\bar\theta_+\gamma^i\dot \theta_- +\frac14\,\bar\theta_-\gamma^i\dot \theta_+\big)
                  \,\big(1-\frac{1}{8R}\,\bar\theta_+\theta_+\big)
          -\frac{\dot t}{4R}\,\big(3\bar\theta_+\gamma^{0i}\theta_-
                                       -\frac{x^k\varepsilon_{ki}}{2R}\,\bar\theta_+\theta_+\big) \,.
\end{align}
The $\kappa$-transformations read as follows:
\begin{align}\label{kappaNH2}
 \delta t &= \frac14\,\bar\kappa\gamma^0\theta_+ \,, \hskip2.45truecm
         \delta x^i = \frac14\,\bar\kappa\gamma^i\theta_- -\frac{\pi^j}{8\pi^0}\,\bar\kappa\gamma_j\gamma_{i0}\theta_+\,, \\
 \delta \theta_+ &=\kappa\,\big(1-\frac{1}{16R}\,\bar\theta_+\theta_+\big) \,, \qquad
         \delta \theta_-=-\frac{\pi^i}{2\pi^0}\,\gamma_{i0}\kappa
              -\frac{3}{8R}\,\gamma_0\theta_-\bar\theta_+\gamma^0\kappa
              +\frac{x^i}{16R^2}\,\gamma^i\kappa\,\bar\theta_+\theta_+\,. \nonumber
\end{align}
Like in the case
of the $\kappa$-symmetric Galilean superparticle discussed above  the $\kappa$-symmetry can be gauge-fixed by
imposing the gauge condition $\theta_+=0$.

%%%%%%%%%%%%%%%%%%%%%%%%%%%%%%%%%%%%%%%%%%%%%%%%%%%%%%%%%%%%%%%%%%%%%%%%%%%%%%%%%%%

\section{Discussion}

In this paper we have constructed the superparticle actions describing the dynamics of a supersymmetric
particle in a 3D Curved Galilean and Newton--Cartan supergravity background. Furthermore, we constructed the
actions for a superparticle moving in the cosmological extension of these backgrounds by including a cosmological
constant. Due to the computational complexity we gave the action in the Newton--Cartan case only up to terms
quartic in the fermions. The Newton--Cartan  background is characterized by more fields and corresponds to more symmetries
than the Galilean background. One can switch between the two backgrounds either by a partial gauging of symmetries
(from Galilean to Newton--Cartan) or by gauge-fixing some of the symmetries (from Newton--Cartan to Galilean). An
important role in the construction is played by symmetries. At several occasions we stressed that, as far as the
background fields are concerned, one should use the background transformations and not the proper
transformations. The latter  are used in the definition of the supergravity multiplet. The relation between
the two kind of transformations is given in eq.~\eqref{rel}.

A noteworthy feature is that the proof of invariance of the superparticle action requires that the background fields
satisfy their equations of motion. This is reminiscent to what happens with the fermionic $\kappa$-symmetry in the
relativistic case. We showed in two particular cases that the non-relativistic superparticle also allows
a $\kappa$-symmetric formulation but that in the non-relativistic case the $\kappa$-symmetry is of a simple St\"uckelberg
type \cite{Gomis:2004pw}. Although being rather trivial, we expect that the formulation with $\kappa$-symmetry is indispensable for a
reformulation of our results in terms of a non-relativistic superspace and superfields, see
e.g.~\cite{Lukierski:2006tr}. Such a superspace formulation would be useful to construct the superparticle actions
in the Newton--Cartan background to all orders in the fermions.

There are several interesting directions in which one could extend our results.
First of all, our efforts in this paper were limited to the three-dimensional case.
Clearly, it would be desirable to construct the four-dimensional analog of our results.
In order to do this, one should first be able to construct the Galilean and Newton--Cartan supergravity multiplets
in four spacetime dimensions. So far, this has not yet been achieved. Another generalization of our results would be to
go from superparticles to superstrings
or even super $p$-branes. This would require taking a `stringy' generalization of the non-relativistic limits we have been
considering here, see e.g.~\cite{Andringa:2012uz}. The case of a non-relativistic superstring in a flat background was
already considered in \cite{Gomis:2005pg}. In the case of a non-relativistic curved background one could apply holography
and study the corresponding non-relativistic supersymmetric boundary theory.

In the case of a (super-)particle propagating in three spacetime dimensions the Galilei algebra contains
a second central charge that could be included as well. This leads to the notion of a {\sl non-commutative}
non-relativistic (super-)particle where the embedding coordinates are non-commutative with respect to the Dirac brackets
\cite{Lukierski:1996br,Batlle:2013yya}.
Finally, we have found that some of our superparticles are 1/2 BPS, corroborating recent results for relativistic
superparticles \cite{Mezincescu:2014zba}. It would be interesting to verify whether the statement that ``all
superparticles are BPS'' \cite{Mezincescu:2014zba} applies to non-relativistic superparticles as well.

%%%%%%%%%%%%%%%%%%%%%%%%%%%%%%%%%%%%%%%%%%%%%%%%%%%%%%%%%%%%%%%%%%%%%%%%%%%%%%%%%%%

\section*{Acknowledgements}

We thank Jelle Hartong and Kiyoshi Kamimura for inspiring discussions and  comments on the draft.
Two of us (E.B. and J.R.) would like to thank the organizers of the {\it Simons Summer Workshop 2013 on Mathematics and Physics}
for its hospitality and generous financial support.
J.G. acknowledges the hospitality at the Department of Theoretical Physics of the University of Groningen where
part of this work was done.
J.G. also acknowledges partial financial support from  the Dutch research organization FOM and from
FPA 2010-20807, 2009 SGR502, CPAN, Consolider CSD 2007-0042.
The work of M.K.~is supported by the Ubbo Emmius Programme administered by the Graduate School of Science, University
of Groningen.
L.P.~acknowledges support by the Consejo Nacional de Ciencia y Tecnolog\'ia (CONACyT), the Universidad Nacional
Aut\'onoma de M\'exico via the project UNAM-PAPIIT IN109013 and an Ubbo Emmius sandwich scholarship from the University of
Groningen.
The work of J.R.~was supported by the START project Y 435-N16 of the Austrian Science Fund (FWF).
T.Z.~acknowledges support by a grant of the Dutch Academy of Sciences (KNAW).

%%%%%%%%%%%%%%%%%%%%%%%%%%%%%%%%%%%%%%%%%%%%%%%%%%%%%%%%%%%%%%%%%%%%%%%%%%%%%%%%%%%

\appendix

\section{The Newton--Hooke Superalgebra}\label{app:A}

The NH (super-)algebra can be derived as a contraction of the AdS (super-)algebra.
In the  case of two supersymmetries there are two independent versions of the latter one, the so-called
$\mathcal{N}=(1,1)$ and $\mathcal{N}=(2,0)$ algebras. In the main text we use only the $\mathcal{N}=(2,0)$
algebra for reasons we will explain below.

We proceed by discussing the contraction of the 3D $\mathcal{N}=(2,0)$ AdS algebra. The basic commutators are given by ($A=0,1,2$)
\begin{align}\begin{split}\label{n20-ads}
 \big[M_{AB},M_{CD}\big] &=2\eta_{A[C}M_{D]B}-2\eta_{B[C}M_{D]A} \,, \qquad     \big[M_{AB},Q^i\big]=-\frac12 \gamma_{AB}Q^i \,, \\
 \big[M_{AB},P_C\big] &=-2\eta_{C[A}P_{B]} \,, \hskip3.6truecm     \big[P_A, Q^i\big]=x\gamma_A Q^i \,, \\[,1truecm]
 \big[P_A,P_B\big] &=4x^2 M_{AB} \,, \hskip4.15truecm    \big[\mathcal{R}, Q^i\big]=2x \,\varepsilon^{ij}Q^j \,, \\[,1truecm]
  \big\{Q_{\alpha}^i,Q_{\beta}^j\big\} &=  2[\gamma^A C^{-1}]_{\alpha\beta} P_A \,\delta^{ij}
            +2x[\gamma^{AB}C^{-1}]_{\alpha\beta}    M_{AB}\,\delta^{ij} +2C^{-1}_{\alpha\beta}\,\varepsilon^{ij}\mathcal{R} \,.
\end{split}\end{align}
Here $P_A, M_{AB}\,, \mathcal{R}$ and $Q_\alpha^i$ are the generators of spacetime translations, Lorentz rotations,
SO(2) R-symmetry transformations and supersymmetry transformations, respectively.
The bosonic generators $P_A$, $M_{AB}$ and $\mathcal{R}$ are anti-hermitian while the fermionic generators $Q_\alpha^i$ are hermitian.
The parameter $x$ is a contraction parameter. Note that the generator of the SO(2) R-symmetry becomes the central
element of the Poincar\'e algebra in the flat limit $x\to0$.

To show that the above  algebra corresponds to the $\mathcal{N}=(2,0)$ AdS algebra it is convenient to define the new generators
\begin{align}
 M_C=\epsilon_{CAB} M^{AB} \,, \hskip1.5truecm J^\pm_A = P_A \pm x M_A \,.
\end{align}
In terms of these new generators we obtain the following (anti-)commutation relations:
\begin{align}
 \big[J^+_A,Q^i\big]=2 x\gamma_AQ^i \,, \hskip1.5truecm
             \big\{ Q^i_\alpha,Q^j_\beta \big\}=2[\gamma^AC^{-1}]_{\alpha\beta} J^+_A \delta^{ij}\,,
\end{align}
while the charges $Q^i$ do not transform under $J^-_A$. This identifies the algebra as the $\mathcal{N}=(2,0)$
AdS algebra.

To make the non-relativistic contraction we define new supersymmetry charges by
\begin{align}\label{Q+Q-}
 Q^\pm_\alpha = \frac{1}{2}\,\big(Q^1_\alpha \pm \gamma_0 Q^2_\alpha \big)\,,
\end{align}
and rescale the generators with a parameter $\omega$ as follows:
\begin{align}\begin{split}\label{contraction}
 P_0 &= \omega\, Z+\frac{1}{2\omega}\,H \,,\qquad \mathcal{R}= -\omega\, Z +\frac{1}{2\omega}\, H \,,
 \qquad M_{a0}= \omega\, G_a \,,\\
 Q^+ &= \frac{1}{\sqrt{\omega}}\, \tilde Q^+ \,, \hskip1.3truecm Q^- = \sqrt{\omega}\,\tilde Q^- \,.
\end{split}\end{align}
We also set $x =1/(2 \omega R)$. Taking the limit $\omega\to\infty$ and dropping the tildes on the $Q^\pm$ we get the
following 3D $\mathcal{N}=(2,0)$ Newton--Hooke superalgebra:
\begin{align}
 \big[J_{ab},(P/G)_c\big] &= -2\,\delta_{c[a}(P/G)_{b]} \,, \hskip.55truecm \big[H,G_a\big] = P_a \,,
             \hskip2.15truecm \big[H,P_a\big] = -\frac{1}{R^2}\, G_a \,, \nonumber\\
 \big[J_{ab}, Q^\pm\big] &= -\frac12\,\gamma_{ab} Q^\pm \,, \hskip1.2truecm \big[H,Q^+\big]=-\frac{1}{2R}\,\gamma_0 Q^+ \,,
               \hskip.6truecm \big[H,Q^-\big]=\frac{3}{2R}\,\gamma_0 Q^- \,, \nonumber\\
 \big[G_a, Q^+\big] &= -\frac12\,\gamma_{a0} Q^- \,, \hskip1.2truecm [P_a, Q^+] = \frac{1}{2R}\,\gamma_a Q^- \,, \label{n20-nh-}\\
 \big\{Q^+_\alpha,Q^+_\beta\big\} &= [\gamma^0C^{-1}]_{\alpha\beta}\,H
                        +\frac{1}{2R}\,[\gamma^{ab}C^{-1}]_{\alpha\beta}J_{ab} \,, \nonumber\\
 \big\{Q^+_\alpha,Q^-_\beta\big\} &= [\gamma^aC^{-1}]_{\alpha\beta}\,P_a
                        +\frac{1}{R}\,[\gamma^{a0}C^{-1}]_{\alpha\beta}G_a \,. \nonumber
\end{align}
The central extension $Z$ that leads to the Bargmann version of the NH superalgebra occurs in the following
(anti-)commutation relations:
\begin{align}
 \big[P_a,G_b\big] = \delta_{ab}\,Z \,, \hskip1.5truecm \big\{Q^-_\alpha,Q^-_\beta\big\} = 2[\gamma^0C^{-1}]_{\alpha\beta}\,Z \,.
\end{align}

The reason why we do not use the $\mathcal{N}=(1,1)$ AdS algebra for the contraction is essentially the same reason as
why we
are interested in $\mathcal{N}=2$ rather than $\mathcal{N}=1$ algebras.  The authors of \cite{Andringa:2013mma} gauged the
$\mathcal{N}=2$ Galilei superalgebra in order to obtain commutator relations that yield $\{Q,Q\}\sim H$ \emph{and}
$\{Q,Q\}\sim \slashed{P}$, that is, the commutator of two supersymmetries gives time- \emph{and} space-translations.
For $\mathcal{N}=1$ one can have only one of them.
The $\mathcal{N}=(1,1)$ AdS algebra is equal to the direct product OSp$(1|2)\otimes $OSp$(1|2)$. Taking the
non-relativistic contraction thereof amounts to taking the simultaneous contractions of two independent
$\mathcal{N}=1$ algebras. However, we already argued that this cannot lead to a superalgebra of the desired form.

%%%%%%%%%%%%%%%%%%%%%%%%%%%%%%%%%%%%%%%%%%%%%%%%%%%%%%%%%%%%%%%%%%%%

\section{Non-linear Realizations}\label{sec:non-linear_realization}

In this appendix we obain the action and transformation rules for the flat Galilean and Newton--Hooke superparticles by
using the method of non-linear realizations \cite{Coleman:1969sm,Callan:1969sn}.\footnote{For an early application of
this method in a different situation than the one considered in this work, namely to the construction of worldline
actions of conformal and superconformal particles, see \cite{Ivanov:1988vw,Ivanov:1988it}.}
We will derive in the first subsection
the $\kappa$-symmetric action and transformation rules of the Galilean superparticle, see e.g.~\cite{Gomis:2006xw,Gomis:2006wu}.
In the second subsection
we will do the same for the NH superparticle. The normalizations of the (to-be) embedding coordinates that
occur in this appendix differ from those in the main text.

\subsection{The Kappa-symmetric Galilean Superparticle}

The starting point is the $\mathcal{N}=2$ Bargmann superalgebra given in section \ref{sec:superparticle}.
We derive the transformation rules for the coordinates $(t,x^i,s,\theta_-^\alpha,\theta_+^\alpha, k^i)$ using the coset
\begin{align}\label{coset}
 g=e^{Ht}\,e^{P_ix^i}\,e^{Zs}\,e^{Q^-_\alpha\theta_-^\alpha}\,e^{Q^+_\alpha\theta_+^\alpha}\,e^{G_ik^i} \,.
\end{align}
This leads to the Maurer--Cartan form
\begin{align}
 \Omega = g^{-1}dg=H\,L_H +\ldots -\bar Q^- L_- -\bar Q^+ L_+ \,,
\end{align}
with the $L$'s given by
\begin{align}\begin{split}
 L_H &= dt -\frac12\,\bar\theta_+\gamma^0d\theta_+ \,, \\
 L_Z &= ds -\bar\theta_-\gamma^0d\theta_-   +\frac{k^ik^i}{2}\,\big(dt-\frac12\,\bar\theta_+\gamma^0 d\theta_+\big)
             +\big(dx^i-\bar\theta_+\gamma^id\theta_-\big) k^i \,, \\
 L_P^i &= dx^i -\bar\theta_+\gamma^id\theta_- +k^i\big(dt-\frac12\,\bar\theta_+\gamma^0d\theta_+\big)
 \,, \hskip2truecm L_G^i = dk^i \,, \\
 L_J^{ab} &= 0\,, \hskip1.9cm  L_- = d\theta_- -\frac12\,\gamma_{i0}d\theta_+\,k^i \,, \hskip1.9truecm L_+ = d\theta_+ \,.
\end{split}\end{align}
It is convenient to define the line-elements
\begin{align} \label{gallineelem}
 \pi^0=\dot t-\frac12\,\bar\theta_+\gamma^0\dot\theta_+ \,, \qquad \pi^i = \dot x^i -\bar\theta_+\gamma^i\dot\theta_-\,,
\end{align}
which are related to the Maurer--Cartan form via the pull-backs
\begin{align}
 (L_H)^* = \pi^0 \,, \qquad (L_P)^*= \pi^i + k^i\pi^0 \,. \label{pidef}
\end{align}

The action of the  Galilean superparticle is given by the pull-back of all $L$'s that are invariant under rotations, hence
\begin{align}
 S &= a\int (L_H)^* +\int (L_Z)^*
   = \int d\tau \,\left[ -\frac{a}{2}\,\bar\theta_+\gamma^0\dot\theta_+ -\bar\theta_-\gamma^0\dot\theta_-
                         -\frac{\pi^i\pi^i}{2\pi^0} \right] \label{galaction}\,.
\end{align}
Here we replaced the Goldstone field $k^i$ by its equation of motion $k^i=-\pi^i/\pi^0$.
This procedure is know as inverse Higgs mechanism \cite{Ivanov:1975zq}, see also \cite{McArthur:2010zm}.
The bosonic transformations of the embedding coordinates are
\begin{align}\begin{split}
 \delta t &=-\zeta \,, \hskip2.55truecm
 \delta x^i =\lambda^i{}_kx^k -a^i +v^it +\frac{v^k\varepsilon_{ki}}{4}\,\bar\theta_+\theta_+\,, \\
 \delta \theta_+  &= \frac14\lambda^{ab}\gamma_{ab}\theta_+     \,,\hskip1.2truecm
 \delta \theta_-  = \frac14\lambda^{ab}\gamma_{ab}\theta_-     +\frac{v^i}{2}\,\gamma^{i0}\theta_+\,,
\end{split}\end{align}
and the supersymmetry transformations are
\begin{align}
 \delta t= \frac12\,\bar\epsilon_+\gamma^0\theta_+ \,, \hskip1.2truecm
 \delta x^i = \bar\epsilon_+\gamma^i\theta_- \,, \hskip1.2truecm  \delta \theta_\pm = \epsilon_\pm \,.
\end{align}
These transformations leave the action \eqref{galaction} and all $L$'s, in particular
the line-elements \eqref{gallineelem}, invariant.

To derive an action that is invariant under $\kappa$-transformations we need to find a fermionic gauge-transformation
that leaves $L_H$ and/or $L_Z$ invariant.\,\footnote{The case of the $SU(1,1|2)$ superconformal particle is discussed
in \cite{Anabalon:2006ii}.}
The variation of $L_H$ and $L_Z$  under gauge-transformations is given by
\begin{align}
 \delta L_H &= d[\delta z_H] -\bar L_+ \gamma^0 [\delta z_+] \,, \\
 \delta L_Z &= d[\delta z_Z] -2\bar L_- \gamma^0 [\delta z_-]
                -\delta_{ab} \big(L_P^a \,[\delta z_G^b] -L_G^a \,[\delta z_P^b] \big) \,.
\end{align}
For $\kappa$-transformations we find, using the explicit expressions for $L_+$ and $L_-$,
\begin{align}
 (\delta L_H)^* &=  [\delta \bar z_+] \gamma^0\dot \theta_+\,, \\
 (\delta L_Z)^* &= 2[\delta \bar z_-]\gamma^0\big(\dot\theta_- -\frac12\,\gamma_{i0}\dot\theta_+\, k^i\big)\,.
\end{align}
It follows that to obtain a $\kappa$-symmetric action we need to take  the pull-back of either $L_H$
or $L_Z$, with $[\delta z_+]=0$ or $[\delta z_-]=0$, respectively. We focus here on the second case, i.e.~we choose $a=0$.
Then the action and $\kappa$-symmetry rules are given by
\begin{align}
 S=\int (L_Z)^* \,, \hskip1.2truecm [\delta z_+]=\kappa \,, \hskip1.2truecm [\delta z_-]=0 \,,
\end{align}
where $\kappa$ is an arbitrary (local) parameter. Using this we find the following $\kappa$-transformations
of the coordinates:
\begin{align}
 \delta t= -\frac12\,\bar\kappa\gamma^0\theta_+ \,, \quad \delta x^i = -\frac{\pi^j}{2\pi^0}\,\bar\theta_+\gamma^i\gamma_{j0}\kappa\,,
 \quad \delta \theta_+=\kappa \,, \quad \delta \theta_-=-\frac{\pi^i}{2\pi^0}\,\gamma_{i0}\kappa \,.
\end{align}
The corresponding $\kappa$-symmetric action  is given by
\begin{align}\label{galkappaaction}
  S = \int d\tau \,\frac{m}{2}\,\left[ -2\bar\theta_-\gamma^0\dot\theta_- -\frac{\pi^i\pi^i}{\pi^0} \right]  \,.
\end{align}

To compare to the action and transformations rules given in the main text one needs to make the following redefinitions
\begin{align}\label{redefinitions}
 t\to -t\,, \qquad x^i\to -x^i+\frac12\,\bar\theta_+\gamma^i\theta_- \,, \qquad \pi^0\to-\pi^0\,,\qquad\pi^i\to-\pi^i\,,
\end{align}
and rescale all spinors by $1/\sqrt2$.

\subsection{The Kappa-symmetric  Newton--Hooke Superparticle}

The previous subsection was  a warming up exercise for the derivation of the  NH superparticle
which is a deformation of the  Galilean superparticle. Here, we give only the main results. Starting from the
NH superalgebra given in appendix \ref{app:A} we choose the same coset as in the previous subsection,
see eq.~\eqref{coset}, and find the following expressions for the $L$'s:
\begin{align}\begin{split}
  L_H &= dt \,\big(1-\frac{1}{4R}\,\bar\theta_+\theta_+\big) -\frac{1}{2}\,\bar\theta_+\gamma^0 d\theta_+ \,, \\
  L_P^i &= dx^i \,\big(1+\frac{1}{4R}\,\bar\theta_+\theta_+\big)
        +\frac{dt}{2R}\,\big(3\bar\theta_+\gamma^{0i}\theta_- -\frac{x^k\epsilon_{ki}}{2R}\,\bar\theta_+\theta_+ \big)
        -\bar\theta_+\gamma^i d\theta_-  +k^i\,L_H \,, \\
  L_Z &= ds +\frac{dt}{2R}\,\big(\frac{x^ix^i}{R}  +3\,\bar\theta_-\theta_-\big) -\bar\theta_-\gamma^0 d\theta_-
        +k^i\,L_P^i -\frac{k^ik^i}{2}\,L_H \,, \\
  L_G^i &= \frac{dt}{R}\,\big(-\frac{x^i}{R} -\frac{x^i}{4R^2}\,\bar\theta_+\theta_+ +\frac{3}{2R}\,\bar\theta_+\gamma^i\theta_-\big)
         -\frac{dx^k\epsilon_{ki}}{4R^2}\,\bar\theta_+\theta_+  -\frac{1}{R}\,\bar\theta_+\gamma^{i0}d\theta_- \\
 &\qquad  -2k^k\,L_J^{ki} +dk^i  \,, \\
  L_J^{ab} &= -\frac{dt\, \epsilon_{ab}}{8R^2}\,\bar\theta_+\theta_+    -\frac{1}{4R}\,\bar\theta_+ \gamma^{ab} d\theta_+\,, \\
  L_- &= d\theta_-\,\big(1-\frac{1}{2R}\bar\theta_+\theta_+\big)
        -\frac{dt}{2R}\,\big(3\gamma_0\theta_- -\frac{3}{2R}\gamma_0\theta_-\bar\theta_+\theta_+
        -\frac{x^i}{R}\,\gamma^{i0}\theta_+\big)         -\frac{dx^i}{2R}\,\gamma^i\theta_+ \\
 &\qquad -\frac{k^i}{2}\,\gamma_{i0}\,L_+ \\
  L_+ &= d\theta_+   +\frac{dt}{2R}\,\gamma_0\theta_+ \,.
\end{split}\end{align}
As before we use $L_P^i$ and $L_H$ to define line-elements using the definition \eqref{pidef}.
These will be useful when we write the action.

Using the Maurer--Cartan form one can derive the transformation rules of the Goldstone fields that realize the
NH superalgebra \eqref{n20-nh-}. We find that the bosonic transformation rules are given by
\begin{align}\begin{split}
 \delta t &=-\zeta \,, \\
 \delta x^i &=\lambda^i{}_kx^k -a^i \cos\frac{t}{R} +\frac{a^k\varepsilon_{ki}}{4R}\sin\frac{t}{R}\bar\theta_+\theta_+
               +v^iR \sin\frac{t}{R} +\frac{v^k\varepsilon_{ki}}{4}\cos\frac{t}{R}\bar\theta_+\theta_+ \,, \\
%  \delta k^i &= \lambda^i{}_kk^k -\frac{a^i}{R}\,\sin\frac{t}{R}\,\big(1-\frac{1}{4R}\,\bar\theta_+\theta_+\big)
%               -v^i\cos\frac{t}{R}\,\big(1-\frac{1}{4R}\,\bar\theta_+\theta_+\big)\,, \\
%  \delta s &= \frac{a^ix^i}{R}\sin\frac{t}{R} +\frac{a^i}{2R}\sin\frac{t}{R}\bar\theta_-\gamma^i\theta_+
%                +v^ix^i\cos\frac{t}{R}  +\frac{v^i}{2}\cos\frac{t}{R}\bar\theta_-\gamma^i\theta_+ \,,\\
 \delta \theta_+  &= \frac14\lambda^{ab}\gamma_{ab}\theta_+     \,,\\
 \delta \theta_-  &= \frac14\lambda^{ab}\gamma_{ab}\theta_-     +\frac{a^i}{2R} \sin\frac{t}{R} \, \gamma^{i0}\theta_+
               +\frac{v^i}{2}\cos\frac{t}{R} \,\gamma^{i0}\theta_+\,.
\end{split}\end{align}
We do not give the transformation rules for $k^i$ and $s$
%guess you found them :)
since we do not need them in the following.
The transformation rules under the $\epsilon_-$-supersymmetry transformations are given by
\begin{align}
 \delta \theta_-=\epsilon_-(t) = {\rm exp}\Big(\frac{3t}{2R}\,\gamma_0\Big)\epsilon_- \,,
%    \qquad \delta s = \bar\epsilon_-(t)\gamma^0\theta_- \,,
\end{align}
while all others fields are invariant (except $s$). This transformation leaves the line-elements invariant. For the
$\epsilon_+$-transformations we find the following rules
\begin{align}\begin{split}
 \delta t&=\frac12\,\bar\epsilon_+(t)\gamma^0\theta_+ \,, \hskip2truecm
         \delta x^i= \bar\epsilon_+(t)\gamma^i\theta_-\,\big(1-\frac{1}{4R}\,\bar\theta_+\theta_+\big) \,, \\
%  \delta k^i &= \frac{k^k}{2R}\,\bar\theta_+\gamma_{ki}\epsilon_+(t) +\frac{x^i}{2R^2}\,\bar\epsilon_+(t)\gamma^0\theta_+
%               -\frac1R\,\bar\theta_-\gamma^{i0}\epsilon_+(t)\,\big(1-\frac{1}{4R}\,\bar\theta_+\theta_+\big)\,, \\
%  \delta s &=   -\frac{x^i}{2R}\,\bar\epsilon_+(t)\gamma^{i0}\theta_-\,\big(1-\frac{1}{4R}\bar\theta_+\theta_+\big)
%                -\frac{x^ix^i}{4R^2}\,\bar\epsilon_+(t)\gamma^0\theta_+
%                -\frac{1}{2R}\,\bar\epsilon_+(t)\gamma^0\theta_+\,\bar\theta_-\theta_- \,,  \\
 \delta \theta_+ &= \epsilon_+(t)\,\big(1-\frac{1}{8R}\,\bar\theta_+\theta_+\big) \,, \\
 \delta \theta_- &= \frac{x^i}{2R}\,\gamma_i\epsilon_+(t)\,\big(1+\frac{1}{4R}\,\bar\theta_+\theta_+\big)
              -\frac{1}{2R}\,\gamma_i\theta_+\bar\epsilon_+(t)\gamma^i\theta_-
              +\frac{3}{4R}\,\gamma_0\theta_-\bar\epsilon_+(t)\gamma^0\theta_+ \,,
\end{split}\end{align}
with
\begin{align}
 \epsilon_+(t) = {\rm exp}\Big(\frac{-t}{2R}\,\gamma_0\Big)\epsilon_+ \,.
\end{align}

We are now ready  to derive the action and $\kappa$-transformation rules.
Like in the Galilean case we need to require that the action consists either of the pull-back of $L_H$ or $L_Z$
but not both, together with $[\delta z_+]=0$ or $[\delta z_-]=0$, respectively. We focus again on the second case,
i.e.~we take
\begin{align}
 S=\int (L_Z)^* \,, \hskip1.2truecm [\delta z_+]=\kappa \,, \hskip1.2truecm [\delta z_-]=0 \,,
\end{align}
with $\kappa$ an arbitrary (local) parameter. Together with the rescalings \eqref{redefinitions} and $R\to-R$ this leads to the
action and transformation rules of the $\kappa$-symmetric NH superparticle
given in section \ref{sec:kappa}, see eqs.~\eqref{kappaNH1} and \eqref{kappaNH2}. After fixing the $\kappa$-symmetry by
imposing the condition $\theta_+=0$ these formulas reduce to those of the  NH superparticle
given in subsection \ref{sec:superNH}.

\section{The NC NH Superparticle: some useful formulas}\label{app:NCNHparticle}

In this appendix we collect a few formulas that are needed to show that the NC NH superparticle action
\eqref{NHparticle} is invariant under the bosonic and fermionic symmetries given in subsection \ref{subsec:superNHNC}.

The transformation rules of the background fields of NC NH supergravity are given by those of NC supergravity
plus the $1/R$ corrections denoted in eqs.~\eqref{NHdelpsi+} and \eqref{NHdelpsi-}. All constraints and equations of
motion on the background fields follow from curvatures of those fields. These curvatures can be derived directly
from the NH superalgebra \eqref{n20-nh-} (with an obvious redefinition of $m_\mu$):
\begin{align}\begin{split}\label{NHcurvatures}
 \hat R_{\mu\nu}(H) &= 2\partial_{[\mu}\tau_{\nu]} -\frac12\,\bar\psi_{+[\mu}\gamma^0\psi_{\nu]+} \,, \\
 \hat R_{\mu\nu}{}^a(P) &= 2\partial_{[\mu}e_{\nu]}{}^a -2\omega_{[\mu}{}^{ab}e_{\nu]b} -2\omega_{[\mu}{}^a\tau_{\nu]}
                        -\bar\psi_{+[\mu}\gamma^a\psi_{\nu]-} \,, \\
 \hat R_{\mu\nu}{}^a(G) &= 2\partial_{[\mu}\omega_{\nu]}{}^a -2\omega_{[\mu}{}^{ab}\omega_{\nu]b}
                        +\frac{2}{R^2}\,e_{[\mu}{}^a\tau_{\nu]} -\frac1R\,\bar\psi_{[\mu-}\gamma^{a0}\psi_{\nu]+} \,, \\
 \hat R_{\mu\nu}{}^{ab}(J) &= 2\partial_{[\mu}\omega_{\nu]}{}^{ab} +\frac{1}{2R}\,\bar\psi_{\mu+}\gamma^{ab}\psi_{\nu+} \,, \\[.1cm]
 \hat R_{\mu\nu}(Z) &= 2\partial_{[\mu} \left( m_{\nu]} +\tau_{\nu]}\,\frac{(x^\rho e_\rho{}^a)^2}{2R^2}\right)
                      -2\omega_{[\mu}{}^ae_{\nu]a} -\bar\psi_{\mu-}\gamma^0\psi_{\nu-} \,, \\[.1cm]
 \hat\psi_{\mu\nu+} &= 2\partial_{[\mu}\psi_{\nu]+} -\frac12\,\omega_{[\mu}{}^{ab}\gamma_{ab}\psi_{\nu]}
                      +\frac1R\,\tau_{[\mu}\,\gamma_0\psi_{\nu]+} \,, \end{split}\\
 \hat\psi_{\mu\nu-} &= 2\partial_{[\mu}\psi_{\nu]-} -\frac12\,\omega_{[\mu}{}^{ab}\gamma_{ab}\psi_{\nu]-}
                      +\omega_{[\mu}{}^a\gamma_{a0}\psi_{\nu]+} -\frac3R\,\tau_{[\mu}\gamma_0\psi_{\nu]-}
                      -\frac1R\,e_{[\mu}{}^a\gamma_a\psi_{\nu]+} \,. \nonumber
\end{align}
The curvatures $\hat R_{\mu\nu}{}^a(P)$ and $\hat R_{\mu\nu}(Z)$ do not change w.r.t.~the NC case discussed in \cite{Andringa:2013mma}.
Therefore one can still use these, by setting $\hat R_{\mu\nu}{}^a(P)=\hat R_{\mu\nu}(Z)=0$, to solve for
$\omega_\mu{}^{ab}$ and $\omega_\mu{}^a$ in terms of the other fields.
Furthermore one can impose the additional constraints
\begin{align}\label{addconstraints}
 \hat R_{\mu\nu}(H)=\hat\psi_{\mu\nu+}=\hat R_{\mu\nu}{}^{ab}(J)=0 \,.
\end{align}
In order to obtain on-shell closure of the supersymmetry algebra on the background fields the curvature
$\hat\psi_{\mu\nu-}$ needs to obey
\begin{align}\label{psi-eom}
 \gamma^\mu\tau^\nu\hat\psi_{\mu\nu-}=0 \,, \hskip2cm e^\mu{}_a e^\nu{}_b \hat\psi_{\mu\nu-}=0 \,.
\end{align}
The additional constraints \eqref{addconstraints} and the equations of motion \eqref{psi-eom} are needed to
proof invariance of
the NC NH superparticle action \eqref{NHparticle}.

{\small

\providecommand{\href}[2]{#2}\begingroup\raggedright\endgroup

% \bibliographystyle{utphys}
% \bibliography{Newton-Cartan}

}

\end{document}